\documentclass[aps,pra,superscriptaddress,showpacs,showkeys]{revtex4}
\usepackage{epsfig}
\usepackage{graphicx}
\usepackage{bm}
\usepackage{amssymb}
\newcommand{\zh}{\bm}

\newcommand{\zhr}{{\zh r}}

\newcommand{\zhe}{{\zh e}}

\newcommand{\zhp}{{\zh p}}
\newcommand{\zhk}{{\zh k}}

\newcommand{\zhA}{{\zh A}}
\newcommand{\zhY}{{\zh Y}}

\newcommand{\Eq}[1]{Eq.\ (\ref{#1})}

\begin{document}

\title{ Effects of autoionization in electron loss from helium-like highly charged ions in collisions with photons and fast atomic particles }

\author{ K.\ N.\ Lyashchenko }
\email{laywer92@mail.ru}
 \affiliation{Department of Physics,
             St.\ Petersburg State University, 7/9 Universitetskaya nab.,
             St. Petersburg, 199034, Russia}
\affiliation{ITMO University,
             Kronverkskii ave. 49, 197101,
             Petergof, St.\ Petersburg, Russia}
\author{O.\ Yu.\ Andreev}
\affiliation{Department of Physics,
             St.\ Petersburg State University, 7/9 Universitetskaya nab.,
             St. Petersburg, 199034, Russia}
\author{ A. B. Voitkiv }
\affiliation{ Institute for Theoretical Physics I,
Heinrich-Heine-University of D\"usseldorf, Germany }
\date{\today}

\begin{abstract}

We study theoretically single electron loss from helium-like highly charged ions involving excitation and decay of autoionizing states
of the ion. Electron loss is caused by either photo absorption or the interaction with a fast atomic particle (a bare nucleus, a neutral atom,  an electron). The interactions with the photon field and the fast particles are taken into account in the first order of perturbation theory. Two initial states of the ion are considered: $1s^2$ and
$(1s \, 2s)_{J=0}$. We analyze in detail how the shape of the emission pattern depends on the atomic number $Z_I$ of the ion discussing, in particular, the inter-relation between electron loss via photo absorption and due to the impact of atomic particles in collisions at modest relativistic and extreme relativistic energies.
According to our results, in electron loss from the $1s^2$ state
autoionization may substantially influence
the shape of the emission spectra only up to
$Z_I \approx 35$--$40$. A much more prominent role is played by
autoionization in electron loss from $(1s \, 2s)_{J=0}$
where it not only strongly affect the shape of the emission pattern
but also may substantially increase the total loss cross section.

\end{abstract}


\maketitle

\section{INTRODUCTION}

\par
Quite often the research in the field of atomic physics is focused ether on the detailed studies of the structure of a single atomic particle
(atom, ion, molecule) or on the dynamics of the interaction between such particles during their collision when subtle details in their structure are normally not important.
In the present paper we will consider a process whose proper description necessitates to take into account both subtle details of the internal structure of the atomic system and the interaction of this system with an incident particle. In this process a helium-like highly charged ion (HCI)
is bombarded by a photon or a fast atomic particle
(electron, nucleus or neutral atom) and, as a result of the interaction between the incident particle and the electrons of the ion,
one of these electrons is emitted (electron loss). We will especially focus on that part of such a process in which
the energy of the final state of the subsystem consisting
of the residual hydrogen-like ion and the emitted electron,
is close to an energy of a doubly excited (autoionizing) state of the initial helium-like ion. In this case the process of electron loss
can be strongly influenced by the presence of autoionizing states:
indeed, the electron loss can proceed not only directly but
also via excitation to autoionizing states with consequent Auger decay (excitation-autoionization (EA) channel).

\vspace{0.15cm}
\par
The processes of photo and charged particle impact ionization of atoms and ions, in which autoionizing states play a significant role, have been attracting much attention (see e.g. \cite{meyer1997lkn,carlson1967lkn,kheifets1998lkn,dorner1996lkn,
wang2006lkn,dubau1973lkn,altick1968lkn,chi1991lkn,hsiao2008lkn,
wang2014lkn,codling1967lkn,wills1998lkn,landers2009lkn,lindsay1992lkn,chi1994lkn,deshmukh1983lkn,
wang1999lkn,bahl1979lkn,kuchiev1985lkn,madden1969lkn,guillemin2012lkn,
sheinerman2006lkn,kuchiev1994lkn,niehaus1977lkn,schmidt1977lkn,heugel2016lkn} for photo ionization, \cite{madden1965lkn,ormonde1967lkn,raeker1994lkn,falk1981lkn,
peart1973lkn,dolder1976lkn,feeney1972lkn,crandall1979lkn,muller1989lkn,
henry1979lkn,tayal1991lkn,pindzola1982lkn,ballance2011lkn,
griffin1984lkn,borovik2011lkn,chen1993lkn,borovik2013kn,fritzsche2012lkn} for charged particle impact, and also references therein).

In particular, the experimental and theoretical studies of photoionization of such atoms, like He \cite{meyer1997lkn,carlson1967lkn,kheifets1998lkn,dorner1996lkn}, Be \cite{wang2006lkn,dubau1973lkn,altick1968lkn,chi1991lkn,hsiao2008lkn}, Ne \cite{wang2014lkn,codling1967lkn,carlson1967lkn,wills1998lkn,landers2009lkn}, Mg \cite{dubau1973lkn,lindsay1992lkn,chi1994lkn}, Ca \cite{deshmukh1983lkn}, Zn \cite{wang1999lkn}, Se \cite{bahl1979lkn}, Ar \cite{kuchiev1985lkn,carlson1967lkn,madden1969lkn,guillemin2012lkn} and Xe \cite{sheinerman2006lkn,kuchiev1994lkn,niehaus1977lkn,kuchiev1985lkn,schmidt1977lkn}, as well as of photodetachment from H$^-$ \cite{meyer1997lkn,kheifets1998lkn}, and of photo electron loss from Li$^+$ \cite{kheifets1998lkn} and Yb$^{2+}$ \cite{heugel2016lkn} have shown
importance of the EA channel and interference between the EA and the direct ionization (detachment, electron loss).

The role of autoionization in collisions of atoms and ions with electrons has been also under a scrutiny. Here, the systems considered include, for instance, helium and helium-like lithium
(see e.g. \cite{madden1965lkn,ormonde1967lkn,raeker1994lkn},
\cite{muller1989lkn}), various ions: Mg$^{+}$, Ca$^{+}$, Sr$^{+}$ \cite{dolder1976lkn}, Ga$^{+}$ \cite{pindzola1982lkn}, Ba$^{+}$
\cite{dolder1976lkn,peart1973lkn,feeney1972lkn}, C$^{3+}$, N$^{3+}$, O$^{3+}$ \cite{crandall1979lkn}, Ti$^{3+}$, Zr$^{3+}$, and Hf$^{3+}$ \cite{falk1981lkn}, Sn$^{q+}$ ($q=4$-$13$) \cite{borovik2013kn}, Xe$^{q+}$ ($q=1$-$6$,$10$) \cite{griffin1984lkn,borovik2011lkn},
light lithium-like ions ranging from B$^{2+}$ to F$^{6+}$ \cite{tayal1991lkn}, and excited C$^{+}$ ions \cite{ballance2011lkn}.
It was, in particular, found that the EA channel and its interference with the direct channel can be of great importance in the processes of electron impact ionization and electron loss.

\par
New features arise in electron loss process when it occurs from highly charged ions where relativistic and QED effects may become of importance. Calculations for electron loss from lithium-like highly charged Ar$^{15+}$, Fe$^{23+}$, Kr$^{33+}$ and Xe$^{51+}$ ions by electron impact were performed in \cite{chen1993lkn}. According to the results reported in \cite{chen1993lkn},  the Breit interaction as well as the M2 transitions may substantially influence the EA channel and the total electron loss. The cross section for electron loss from beryllium-like HCIs in collisions with bare atomic nuclei, differential in
the emission angle, was calculated in \cite{fritzsche2012lkn}.
The authors of \cite{fritzsche2012lkn} predict
a significant impact of the Breit interaction
on the EA channel involving the $1s2s^22p_{3/2} \phantom{1}^3P_2$ state.

\par
A nice review on the various aspects of autoionization
in electron-ion collisions can be found in \cite{muller2008lkn_book}.

\vspace{0.15cm}

\par
In the present paper we study single electron loss
from helium-like HCIs by photo absorption and by the impact
of fast atomic particles. In this consideration we will assume
that the interactions of the HCI with the photon field involves
only one photon from this field. We shall also suppose that the interaction between the atomic particle and the electrons
of the HCI is sufficiently weak such that it can be described within the first order of relativistic perturbation theory.
Thus, both these electron loss processes involve the exchange
of just one photon, real or virtual, between the colliding systems.

The main interest of the present study concerns the role of autoionization in the loss processes which can be significant when the energy of the emitted electron is close to
the difference between the energy of an autoionizing state of a helium like HCI and the energy of the final state of the residual hydrogen-like ion. In such a case the autoionizing state can participate in the process via the corresponding EA channel. This is illustrated
in Fig. \ref{fig1}, where both the direct and indirect (EA)
channels of electron loss are schematically depicted.

\par
Two initial states of the HCI will be considered:
the ground $(1s^2)$ state and the metastable $(1s \, 2s)_{J=0}$ state.
In both cases the residual  hydrogen-like ion is supposed
to be in the ground state. Note that a transition from the ground state of a helium-like ion
to its autoionizing state via a single interaction with an external
field becomes possible only due to the interaction (correlation)
between the electrons of the ion. Therefore, electron loss
from the ground state via excitation and decay of autoionizing states represents a highly correlated process.

\par
The paper is organized as follows.
In the next section we consider theoretical
treatments for electron loss via photo absorption
and by the impact of atomic particles
and obtain cross sections for these processes.
Results of our numerical calculations and their discussion
are presented in section III. Section IV contains summary.

The relativistic units ($\hbar = c = m_e = 1$)
are used throughout unless otherwise stated.

\section{GENERAL CONSIDERATION}

Our description of electron loss from helium-like HCIs by photo
absorption and by the impact of fast atomic particles is based on
the line-profile approach (LPA) \cite{andreev2009lkn,andreev2008lkn} of QED. The Furry picture is used, in which the Coulomb interaction of the electrons of the HCI with its nucleus is fully taken into account from the onset. The Dirac equation is employed to treat both bound electrons
and the emitted electron. The interaction of these electrons
with the quantized electromagnetic and electron-positron fields is  treated according to the standard QED perturbation theory. This interaction is taken into account in the zeroth and first orders of the perturbation expansion. Besides, the leading parts of the higher order corrections in the electron-electron interaction are also included into the treatment according to the LPA \cite{andreev2008lkn}.

\par
The interaction between the electrons of the HCI with the field of an  incident fast atomic particle (an electron, a nucleus, a neutral atom) is described within the first order of relativistic perturbation theory. This implies that the field of this particle has to be relatively weak, which is the case if the conditions $\frac{e^2}{ \hbar v }\ll 1$ or $\frac{Z_a e^2}{\hbar v} \ll 1$, where $|e|$ is absolute charge of the electron, $Z_a$ is the charge of the incident nucleus (or the atomic number of the incident atom) and $v$ is the collision velocity.

\par
In Fig. \ref{fig2} we illustrate our description of electron loss
from the ground state by the impact of an atomic particle by presenting Feynman graphs which  correspond to the zeroth and first orders of the perturbation theory with respect to the interaction between the electrons of the HCI. The graph \textit{a} represents the main contribution to the direct channel of electron loss in which one of the electrons of the ion, via absorbtion of a virtual photon, makes a transition to the continuum, whereas the other electron remains merely a spectator. This graph obviously does not contribute to the indirect channel (EA) of electron loss, in which both electrons are excited into an autoionizing state with its consequent
Auger decay. In contrast, the graphs \textit{b} - \textit{e} contribute
both to the direct and indirect channels.

\subsection{ Construction of two-electron states of a free HCI }

\par
In the framework of the LPA the initial and final states of a free
HCI in the zeroth order of perturbation theory are chosen as configurations of noninteracting electrons in the $j$-$j$ coupling scheme. The interaction with the quantized fields leads to various corrections to these states.
They are given as linear combinations of many different configurations
of the noninteracting electrons in the $j$-$j$ scheme. The coefficients
in these linear combinations (the mixing coefficients) are obtained
in the framework of the LPA.

\par
Employing the rest frame of the HCI, taking the position of the HCI's nucleus as the origin and denoting by ${\bf r}_1$ and ${\bf r}_2$ the coordinates of the electrons of the HCI, the two-electron wave functions in the zeroth order of perturbation theory are given by
\begin{eqnarray}
\Psi_{JMn_1j_1l_1n_2j_2l_2}^{(0)}({\bf{r}}_1,{\bf{r}}_2)
&=&\nonumber
N\sum_{m_1,m_2}C_{JM}^{j_1,j_2}(m_1,m_2)\\
&\times&
\det\{\psi_{n_1j_1l_1m_1}({\bf{r}}_1)\psi_{n_2j_2l_2m_2}({\bf{r}}_2)\}
\,.
\end{eqnarray}
Here, $N$ is the normalization constant (equal to $1/2$ for equivalent electrons and to $1/\sqrt{2}$ otherwise), $C_{JM}^{j_1,j_2}(m_1,m_2)$
are the Clebsch-Gordan coefficients, $\psi_{n_i,j_i,l_i,m_i}$ is
a solution of the one-electron Dirac equation where $n_i$
is the principal quantum number, $j_i$ is the total angular momentum,
$m_i$ is its projection and $l_i$ defines parity ($(-1)^{l_i}$),
where the index $i$ ($i=1$, $2$) labels the \textit{i}-th electron
of the HCI.

\par
The emitted electron with momentum ${\bf{p}}$, energy $\epsilon_p=\sqrt{1+p^2}$ and polarization $\mu$ is described
by the following expansion \cite{akhiezer1965}
\begin{eqnarray}
\psi_{\zhp, \mu}(\zhr)
&=&\label{psie}
\int d\varepsilon \sum_{jlm} a_{{\zhp} \mu, \varepsilon jlm}\psi_{\varepsilon jlm}(\zhr)
\,,
\end{eqnarray}
where
\begin{eqnarray}
a_{{\zhp} \mu, \varepsilon jlm}
&=&\label{coefa}
\frac{(2\pi)^{3/2}}{\sqrt{p \epsilon_p}}i^l e^{-i\phi_{jl}}
(\Omega^+_{jlm}(\zhp), \upsilon_{\mu}(\zhp))
\delta(\epsilon_p-\varepsilon)
\,.
\end{eqnarray}
In \Eq{coefa} the $\phi_{jl}$ is the Coulomb phase shift,
$\Omega_{jlm}(\zhp)$ is the spherical spinor
and $\upsilon_{\mu}(\zhp)$ is the spinor with
projection $\mu=\pm 1/2$ on the electron momentum ($\zhp$)
\begin{eqnarray}
\frac{\zhp \, \hat{\zh\sigma}}{2p} \upsilon_{\mu}(\zhp)
&=&
\mu
\upsilon_{\mu}(\zhp)
\,,
\end{eqnarray}
where
$\hat{\zh\sigma}$
is the Pauli vector.
It is convenient to describe the final two-electron state ($1s, \, e^{-}$), which contains the emitted electron, as a linear combination of the following determinants
\begin{eqnarray}
\frac{1}{\sqrt{2}}\det\{\psi_{n_bj_bl_bm_b}({\bf{r}}_1)\psi_{\zhp,\mu}({\bf{r}}_2)\}
&=&\label{psi_f}
\sum_{JMjlm}C_{JM}^{j_b,j}(m_b,\mu)
\int d\varepsilon \, a_{{\zhp} \mu, \varepsilon jlm}
\Psi_{JMn_bj_bl_b\varepsilon jl}^{(0)}({\bf{r}}_1,{\bf{r}}_2)
\,,
\end{eqnarray}
where the quantum numbers $n_b$, $j_b$, $l_b$ and $m_b$ refer
to the $1s$-electron in the resudial hydrogen-like HCI.

\par
In order to take into account the QED corrections such as the interelectron interaction correction, the electron self-energy,
and the vacuum polarization corrections we employ the LPA.
According to the LPA we construct new functions which are
given by \cite{andreev2008lkn}
\begin{eqnarray}
\Phi_m(\zhr_1,\zhr_2)
&=&\label{Pni_lpa}
\sum_{k\in g} B_{km}\Psi_k^{(0)}(\zhr_1,\zhr_2)
+\sum_{n \notin g,l\in g} [\Delta V]_{nl}\frac{B_{lm}}{E^{(0)}_{m}-E^{(0)}_{n}}\Psi_n^{(0)}(\zhr_1,\zhr_2)
\,,
\end{eqnarray}
where $m$, $k$, $l$ and $n$ label the two-electron configurations
and $g$ denotes the set of all two-electron configurations considered in this work. The set $g$ includes all the two-electron configurations
composed of $1s$, $2s$, $2p$, $3s$, $3p$, $3d$, $4s$, $4p$, $4d$, $4f$ electrons and the electron in the continuum. The $\Delta V$ represents the matrix of the QED corrections: the one- and two-photon exchange corrections, the electron self-energy correction, and the vacuum polarization correction. The first term on the right hand side in \Eq{Pni_lpa} corresponds to the mixing of the configurations from the set $g$. The mixing coefficients $B_{km}$ are obtained within the LPA. The second term on the right hand side in \Eq{Pni_lpa} takes into account all other configurations which are not included in the set $g$. The two-electron configuration energy in zeroth order $E^{(0)}_{n}$ is the sum of the corresponding one-electron Dirac energies.

\par
For example, the ground two-electron state
constructed according to \Eq{Pni_lpa}
has the following form
\begin{eqnarray}
\Phi_{(1s)^2}
&=&\label{mix_coef}
B_{(1s)^2,(1s)^2}\Psi_{(1s)^2}^{(0)}+B_{(1s2s),(1s)^2}\Psi_{(1s2s)}^{(0)}+ B_{(1s3s),(1s)^2}\Psi_{(1s3s)}^{(0)}
+...
\,,
\end{eqnarray}
where only the main terms of the first sum
in \Eq{Pni_lpa} are explicitly given.
In order to describe the strongest indirect channels
of electron loss from the ground state of a HCI
(which are possible via just a single interaction with an external field)
it is necessary to take into account the contributions from the $(1s 2s)$ and $(1s 3s)$ configurations. The absolute values of the mixing coefficients corresponding to these configurations are shown
in Fig.\ref{fig3} where they are given
as a function of the charge $Z_I$ of the HCI's nucleus.
It is seen in the figure that the initial rapid decrease in
the magnitude of the mixing coefficients at not very large $Z_I$
is then replaced by a much slower decrease at higher $Z_I$ and
that these coefficients become almost constants at very high $Z_I$.

This behaviour can be understood by noting that the
Coulomb and the Breit parts of the electron-electron interaction
scale approximately as $\sim Z_I$ and $Z_I^3$, respectively.
Since the energy differences, $E^{(0)}_{n}$ - $E^{(0)}_{m}$ ($n \neq m$), scale with $Z_I$ as $Z_I^2$, the contribution of the Coulomb and Breit parts to the mixing coefficients are proportional to $1/Z_I$ and $Z_I$, respectively. Therefore, at not very high $Z_I$, where the Coulomb part
of the electron-electron interaction strongly dominates,
we observe that the mixing coefficients depend on $Z_I$ as $1/Z_I$
whereas at very high $Z_I$, where the effective strength of Breit interaction becomes on overall comparable to that of the Coulomb one,
these coefficients are almost $Z_I$-independent.

Within our treatment the transitions between the ground and
an autoionizing state via absorption of just one photon (real or virtual)
is only possible due to the admixture of other configurations
to the main configuration in these states.
Since the strength of this admixture
depends on the absolute values of the mixing coefficients,
one can conclude from Fig.\ref{fig3} that the role of autoionizing states in the electron loss from the ground state rapidly
decreases with increasing $Z_I$. In particular, our calculation
show that beginning with $Z_I \approx 35$--$40$ autoionizing states
do not noticeably influence even differential spectra of
electron loss from the ground state.

\subsection{ Electron loss by photo absorption }

Let us first consider electron loss from a helium-like HCI by absorption of a (real) photon. In Fig. \ref{fig1} both the direct and indirect (EA)  channels of electron loss are depicted. As was already mentioned,
the presence of autoionizing states will influence the process of electron loss if the energy of the emitted electron is close to
$ \epsilon_p = E_a - \varepsilon_{1s}$, where $E_a$ is the energy of an autoionizing state and $\varepsilon_{1s}$ is the energy of the ground state of the residual hydrogen-like ion.

\par
The amplitude of electron loss via photo absorption,
which includes both the direct and indirect channels, reads
\begin{eqnarray}
A_{if}
&=&
- i \langle \Phi_f| \Xi |\Phi_i\rangle
\,,
\end{eqnarray}
where $\Xi$ is the photon absorption operator \cite{andreev2008lkn}.
In general this operator  can be derived within LPA by perturbation expansion. In our case it sufficient to take $\Xi$ in zeroth order form

\begin{eqnarray}
\Xi^{(0)}_{if}
&=&
 e \langle \Phi_f (\zhr_1,\zhr_2)| \gamma^{\nu}A^{(\zhk,\lambda)}_{\nu}(\zhr_1) + \gamma^{\eta}A^{(\zhk,\lambda)}_{\eta}(\zhr_2) |\Phi_i(\zhr_1,\zhr_2) \rangle
\,,
\end{eqnarray}
where $\gamma^{\nu}$ are Dirac gamma matrices. Four vector $A^{(\zhk,\lambda)}_{\nu}=(A_0^{(\zhk,\lambda)}, \zhA^{(\zhk,\lambda)})$ is the emitted photon wave function. We use the transverse gauge in which $A_0^{(\zhk,\lambda)}=0$ and $\zhA^{(\zhk,\lambda)}$ reads
\begin{eqnarray}
{\zhA}^{(\zhk,\lambda)}({\zhr})
&=&
\sqrt{\frac{2\pi}{\omega}} e^{i \zhk\zhr} \zhe^{(\lambda)}
\,,
\end{eqnarray}
where $\omega$, $ \zhk=(0,0,k)$ and $\zhe^{(\lambda)}$ ($ \lambda = 1,2$)  are the frequency, momentum and polarization vector of the photon, respectively.
Employing the multipole expansion we can write
\cite{labzowsky96b}
\begin{eqnarray}
 {\zhA}^{(\zhk,\lambda)} (\zhr)
&=&\label{12345}
\sqrt{\frac{2\pi}{\omega}}
\sum_{j_0l_0m_0}i^{l_0} g_{l_0}(\omega r)
(\zhe^{(\lambda)} \cdot {\zhY}_{j_0l_0m_0}^*(\zhk)) \zhY_{j_0l_0m_0}(\zhr)
\,,
\end{eqnarray}
where $g_{l_0}(x)=4\pi j_{l_0}(x)$ and $j_{l_0}(x)$ is the spherical Bessel function,
$\zhY_{j_0l_0m_0}$ is vector spherical harmonics, and $\zhe^{(\lambda)}$ is vector of photon polarization. 

\par
The cross section differential in the emission angles
reads
\begin{eqnarray}
\frac{d \sigma}{d\Omega}
&=&\frac{1}{2}\label{sigmaph}
\sum_{\lambda}\frac{\epsilon_p p}{(2 \pi )^2} |A_{if}|^2
\,.
\end{eqnarray}


\subsection{ Electron loss in collisions with a charged particle }

Similarly to the case of electron loss via photo absorption,
it is convenient to give the basic consideration of electron
loss by the impact of a fast particle using the rest frame of the HCI.
We again take the position of the HCI's nucleus as the origin and
denote by ${\bf r}_1$ and ${\bf r}_2 $ the coordinates of the electrons of the HCI. We shall suppose that the energy of the incident particle is so high that its change in the process is negligible compared to its initial value. This allows us to describe the particle as moving along a straight-line classical trajectory ${\bf R}(t) = ({\bf{b}}, vt)$, where ${\bf{b}}=(b_x,b_y,0)$ is the impact parameter, ${\bf{v}}=(0,0,v)$ is the velocity of the particle and $t$ denotes the time.

Within the first order of perturbation theory in the interaction between the electrons of the HCI and the field of the incident particle with a charge $Z_p$ the amplitude of the electron loss from the HCI is given by
\begin{eqnarray}
A_{if}({\bf{b}})
&=&\label{A_b}
-i \int^{\infty}_{-\infty} dt \,  \langle \Phi_f| W_{{\bf{b}}}({\bf{r}}_1,t)+ W_{{\bf{b}}}({\bf{r}}_2,t)|\Phi_i\rangle e^{it(E^{(0)}_{f}-E^{(0)}_{i})}
\,,
\end{eqnarray}
\begin{eqnarray}
W_{{\bf{b}}}({\bf{r}_i},t)
&=&\label{W_b}
-\frac{\gamma Z_p e^2 (1-v \alpha_z) }{\sqrt{({\bf{b}}-{\bf{r}}_{i,{\bot}})^2 + \gamma^2 (vt-r_{i,z})^2}} \,,
\end{eqnarray}
where $W_{{\bf{b}}}({\bf{r}}_i,t)$ is the interaction
between the projectile and \textit{i}th ionic electron ($i=1,2$).
In \Eq{W_b} $ \gamma = 1/\sqrt{1 - v^2} $ is
the Lorentz factor of the collision, and $\alpha_z$
is the corresponding Dirac alpha matrix.

\par
It is convenient to work with the transition amplitude
in the momentum space which is obtained from \Eq{A_b} by performing
the Fourier transformation
\begin{eqnarray}
S_{if}({{\bf{q}}_{\bot}})
&=&
\frac{1}{2\pi}\int d^2 {\bf{b}} \, A_{if}({\bf{b}}) e^{i{\bf{q}}_{\bot} {\bf{b}}}
\,,
\end{eqnarray}
which results in
\begin{eqnarray}
S_{if}({{\bf{q}}_{\bot}}) &=& \label{S_q}
2i \frac{Z_p e^2}{v}\frac{\langle \Phi_f|
W_{ {\bf q}_\perp }({\bf{r_1}}) + W_{ {\bf q}_\perp }({\bf{r_2}})
|\Phi_i \rangle}{{ q^{2}_{\bot}} + \frac{q_{min}^2}{\gamma^2} }
\,,
\end{eqnarray}
where
\begin{eqnarray}
W_{{\bf{q}}_{\bot}}({\bf{r}}_i)
&=&\label{ampl_q}
e^{i {\bf{q}} {\bf{r}}_i} (1 - v \alpha_z)
\,.
\end{eqnarray}
In the above expressions the momentum transfer
${\bf q}$ to the HCI in the collision reads
\begin{eqnarray}
{\bf{q}} &=& ({{\bf{q}}_{\bot}},q_{min})
\nonumber \\
q_{min} &=& \frac{E^{(0)}_f-E^{(0)}_i}{v} \, ,
\label{momentum_transfer}
\end{eqnarray}
where $E^{(0)}_i$ is the energy of the initial state of the HCI
and $ E^{(0)}_f$ is the sum of the energies of the emitted electron
and the residual hydrogen-like ion.

\par
The fully differential cross section for electron loss by particle
impact is given by
\begin{eqnarray}
\frac{d^4 \sigma}{d\epsilon_p d\Omega d^2{{\bf{q}}_{\bot}}}
&=&\label{sigma}
\frac{\epsilon_p p}{(2 \pi )^3} |S_{if}({{\bf{q}}_{\bot}})|^2
\,.
\end{eqnarray}

\subsection{ Electron loss in collisions with a neutral atom }

When an ion collides with a neutral atom, the interaction between the electrons of the ion both with the nucleus of the atom and the electrons of the atom can contribute to electron loss from the ion. According to the consideration of first order in the interaction between the ion and atom, the role of the atomic electrons in the electron loss is two-fold.

\par
First, in collisions, in which the atom remains in its initial internal state (elastic atomic mode), the electrons of the atom tend to screen (fully or partially) the charge of the atomic nucleus (see e.g.
\cite{stolterfoht1997lkn_book,mcguire1997lkn_book,voitkiv2008lkn}).
The screening effect reduces the probability for electrons of the ion to undergo a transition compared to collisions with the corresponding bare atomic nucleus. Second, in collisions, which are inelastic also for the atom, only the interaction with the electrons of the atom may contribute to transitions of the ionic electrons (see e.g.
\cite{stolterfoht1997lkn_book,mcguire1997lkn_book,voitkiv2008lkn}). In this case the presence of the atomic electrons increases the probability for the electrons
of the ion to make a transition \cite{antiscreening}.

\par
In collisions, which are characterized by momentum transfers to the atom  substantially exceeding typical orbiting momenta of the atomic electrons, the contributions of the elastic and inelastic atomic modes to the cross sections for excitation of (electron loss from ) the ion roughly scale
as $\sim Z_A^2$ and $\sim Z_A$, respectively,
where $Z_A$ is the atomic number of the atom
(see e.g. \cite{voitkiv2008lkn}).
In the present paper we consider collisions of HCIs
with multielectron atoms ($Z_A \gg 1$) only, where the relative contribution of the inelastic mode is small. Therefore, in what follows
we shall simply neglect the inelastic atomic mode. The contribution from the elastic atomic mode can be obtained from expression \Eq{S_q} if we replace $Z_p$ by the effective charge of the atom given by \cite{voitkiv2008lkn}
\begin{eqnarray}
Z_{A,eff}
&=&
Z_{A} \left( {{q}^2_{\bot}} +  \frac{q_{min}^2}{\gamma^2}  \right) \sum_{i=1}^3 \frac{A_i}{k_i^2+{{q}^2_{\bot}} + \frac{q_{min}^2}{\gamma^2}}
\,,
\label{effect_charge}
\end{eqnarray}
where the parameters $A_i$ and $k_i$ ($i=1,2,3$) are tabulated
for various atoms in \cite{moliere1947lkn,salvat1987lkn}.

\section{ Results and discussion }

In this paper we present results for electron loss
assuming that the incident particles as well as
the residual hydrogen-like ion and the emitted electron
are unpolarized.

\subsection{ Electron loss from the ground state }

The process of electron loss looks most simple when
it is considered in the rest frame of the HCI. Therefore,
we start with presenting results obtained
for this frame.

Results of our calculations for electron loss from
Ca$^{18+}$(1s$^2$) and Zn$^{28+}$(1s$^2$) by photo absorption are shown in panels a) and b) of Figs. \ref{fig4} -- \ref{fig7}. In the rest frame of the HCI the incident photon is supposed to move
along the $z$-axis (from which the polar angle for
the emitted electron is counted). The main feature in the energy spectrum
of the emitted electrons, displayed in Figs. \ref{fig4}a and \ref{fig5}a,
is a number of structures each consisting of groups
of narrow maxima and minima.
These structures arise due to the EA channel and its interference with
the direct channel of electron loss. They correspond to
the participation of $LL$, $LM$ and $MM$ autoionizing states
and are embedded in a smooth background which appears solely due to the direct electron loss channel (see Fig. \ref{fig1}). As one expects
from the dependence of the mixing coefficients on $Z_I$, these structures are more pronounced in electron loss from the lighter ion where the EA channel plays a more prominent role.

These structures are also seen in the energy-angular distribution
of the emitted electrons (Figs. \ref{fig4}b and \ref{fig5}b).
Inspecting this distribution one can note that with increasing
emission energy the maximum in the emission slightly shifts to lower emission angles that is related to the corresponding increase in the momentum of the absorbed photon.

\vspace{0.25cm}

Our results for electron loss from Ca$^{18+}$ and Zn$^{28+}$
in collisions with bare nuclei of nitrogen (N$^{7+}$) are presented
in Figs. \ref{fig4}c,d -- \ref{fig7}c,d for a  modest relativistic
impact energy of $93$ MeV/u ($\gamma = 1.1$, $v \approx 0.42 \, c$).
We again observe familiar structures in the emission pattern caused by the excitation and Auger decay of autoionizing states and interference with the direct electron loss channel.

However, since polarizations and momenta of a real and virtual photon
(both having the same energy) are quite different,
compared to electron loss via photo absorption the emission pattern
in collisions with charged particles has a number of distinct features.
In particular, because of less strict selection rules for
electron transitions in collisions with charged particles,
now more autoionizing states contribute to the loss process.
Besides, the angular distribution of the emitted electrons becomes  noticeably broader and the maximum in the emission is shifted
to smaller emission angles.
Moreover, the differences in the shape of the emission pattern
in electron loss from Ca$^{18+}$ and Zn$^{28+}$ are much more obvious
in collisions with charged particles than in photo absorption.

The change in the shape of the emission spectra, which we observe when
$Z_I$ varies, is caused by the $Z_I$-dependence of the ratio between the typical momentum transfer to the HCI in the collision and the typical orbiting electron momenta in the HCI. Indeed, the former and the latter scale with $Z_I$ approximately
as $\sim Z_I^2$ and $\sim Z_I$, respectively.
The growth of the ratio leads to the increase of the effective
number of the partial waves contributing to the state of
the emitted electron that changes the shape of the emission pattern.
Besides, as calculations show, the effective magnitude
of the transverse component $q_\perp$ of the momentum transfer
${\bf q}$ increases with $Z_I$ somewhat less rapidly than
its longitudinal component $q_{min}$ ($q_{min} \sim Z_I^2$).
This makes the orientation of ${\bf q}$ effectively $Z_I$-dependent.
Since the relative populations of the magnetic sublevels of the autoionizing and continuum states in the collision depend
on the orientation of ${\bf q}$ with respect to the quantization axis,
the relative populations of these sublevels
for Ca$^{18+}$ and Zn$^{28+}$ turn out to be different
that also leads to changes in the emission pattern.

\vspace{0.25cm}

Cross sections for electron loss from Ca$^{18+}$
in collisions with neutral nitrogen \cite{nitrogen}
at an impact energy of $93$ MeV/u are displayed
in Figs. \ref{fig4}e,f and \ref{fig6}e,f.
Comparing them with the corresponding results for
collisions with equivelocity bare nitrogen nuclei
shown in Figs. \ref{fig4}c,d and \ref{fig6}c,d we see
that the presence of atomic electrons has a rather
weak effect on the electron loss process.
The reason for this is that the electrons
in the HCI are tightly bound and, as a result, the
loss process involves momenta transfers which are large
on the typical atomic scale of nitrogen that makes
the effective charge of the atom $Z_A(q)$, given
by Eq.(\ref{effect_charge}), to be approximately equal
to the charge of the unscreened atomic nucleus for
the most important range of the momentum transfers
contributing to the loss process.

Note also that the origin of the weakness of the screening effect
in collisions with HCI can be viewed from a different perspective.
Namely, due to the small size of the ground state of electrons
tightly bound in the HCI (and because of not very large impact velocity)
the loss process takes place mainly in collisions with
impact parameters much smaller than the size
of the atom. Therefore, only a small fraction of the atomic
electron cloud can stay between the electrons of the HCI and
the nucleus of the atom that weakens the screening effect.

Since the electrons in Zn$^{28+}$(1s$^2$) are tighter bound
than those in Ca$^{18+}$(1s$^2$), the screening effect
of the atomic electrons in the electron loss from Zn$^{28+}$
is even weaker. Therefore, results shown in Figs. \ref{fig5}c,d and
\ref{fig7}c,d for electron loss by collisions with N$^{7+}$
essentially represent also results for collisions with neutral
nitrogen.

\vspace{0.25cm}

In Figs. \ref{fig6} and \ref{fig7} we focus on the range of emission energies for which the $(2s \, 2p_{3/2})_{J=1}$ autoionizing state actively participates resulting in the strongest resonance \cite{strongest_resonance} (for more information about the parameters
of the resonance see Table 1). The energy spectra, shown in the left panels of these figures display one of characteristic shapes known
as Fano profiles \cite{fano1961lkn}. Such a shape arises because
of interference between the direct and indirect (EA) channels
of electron loss.

For this quite narrow emission energy range
both photo absorption and the impact of a fast
atomic particle lead to a basically similar shape of
the energy spectra although in the latter case the interference
is less pronounced because more continuum partial waves contribute
to the direct channel of electron loss in collisions with atomic particles than in photo absorption due to larger values of the momentum transfer (at the same energy transfer) and also "softer" selection rules. However, the differences between these two processes become much more obvious
when the energy-angular distributions of the emitted electron
are considered (see the right panels of Figs. \ref{fig6} and \ref{fig7}).

Note also that the contribution of
vacuum polarization and electron self-energy to the position of
the resonance is much larger than the total width
of the autoionizing state (see Table 1) that makes it necessary
to take this contribution into account for a precise determination
of this position.

\subsection{ Electron loss from the initial $(1s \, 2s)_{J=0}$ state }

Comparing results shown in Figs. \ref{fig4} and \ref{fig5}
we may conclude that, as was already expected, the role
of autoionization in the process of electron loss diminishes
with increasing the atomic number of the HCI. For helium-like HCIs
with $Z_I$ substantially higher than those considered in the previous subsection, the role of these states in electron loss from the ground state becomes negligible.

The situation, however, changes drastically if electron loss from
a helium-like HCI occurs from the metastable $(1s \, 2s)_{J=0}$ state.
In such a case autoionizing states with the basic configuration
involving a $2s$-electron can be reached from the metastable
$(1s \, 2s)_{J=0}$ state by essentially a single-electron transition
(i.e. the electron-electron correlations are no longer necessary
for these transitions to occur)
that greatly enhances the role of autoionization
in the electron loss process. Indeed, now a single-electron
transition into an autoionizing state
can more effectively compete with the direct (single-electron)
transition into the continuum since the strength of both
of them scale with $Z_I$ similarly. Therefore, in electron loss
from the $(1s \, 2s)_{J=0}$ state, autoionization plays
a much more prominent role.

The energy and energy-angular distributions of electrons emitted
from helium-like Xe$^{52+}(1s \, 2s)_{J=0}$ and U$^{90+}(1s \, 2s)_{J=0}$ ions due to photo absorption and in collisions with $93$ MeV/u nitrogen
\cite{footnote} are shown in Figs. \ref{fig8} and \ref{fig9}
for a rather broad range of emission energies,
whereas in Figs. \ref{fig10} and \ref{fig11} we again focus
on the narrow range of emission energies for which
the $(2s \, 2p_{3/2})_{J=1}$ autoionizing state
takes active part in the loss process.
It is seen that photo absorption and atomic particle impact
lead in general to qualitatively different emission patterns.

A comparison of the energy and energy-angular distributions of electrons emitted from Ca$^{18+}(1s \, 2s)_{J=0}$,
Xe$^{52+}(1s \, 2s)_{J=0}$ and U$^{90+}(1s \, 2s)_{J=0}$ in collisions with N$^{7+}$ at an impact energy of $93$ MeV/u
is presented in Fig. \ref{fig12} for the range
of emission energies corresponding to the involvement
of the $(2s \, 2p_{3/2})_{J=1}$ autoionizing state.
As it follows from these figures, the autoionizing states play now much
more prominent role. In particular, for electron loss from
Ca$^{18+}$ the resonance in the energy spectrum of the emitted electron
corresponding to the EA channel is so strong
that in collisions with atomic particles
its contribution to the total loss cross section
is comparable to that of the direct channel.

However, we see that the influence of autoionization
still diminishes with increasing $Z_I$. For instance,
whereas autoionizing states continue to have a noticeable
impact on the total cross section for electron loss
from Xe$^{52+}$, the same already cannot be said
in case of U$^{90+}$. Yet, the origin of this decreasing role
of autoionization with increasing $Z_I$ lies
now not in the different $Z_I$-scalings for the transitions
via the EA and direct channel but is solely caused
by the competition between
the Auger decay of autoionizing states and
their spontaneous radiative decay.

For atoms and relatively low charged ions
the Auger decay is known to dominate the spontaneous
radiative decay. However, the rate of the latter
very rapidly grows with $Z_I$ whereas the rate
of the former depends on $Z_I$ relatively weakly.

For electron loss from Ca$^{18+}$,
for which these two rates are rather close,
the EA channel is still so strong that it not
only completely dominates the direct channel
in the vicinity of the resonance (see Fig. \ref{fig12} a,b)
but also gives a very substantial contribution
to the total electron loss.

In the case of Xe$^{52+}$ the spontaneous radiative decay
of the autoionizing state is already much faster
than its Auger decay that makes the EA channel
less efficient (see Fig. \ref{fig12} c,d).
Nevertheless, this channel
still greatly dominates the electron loss process
in the vicinity of the resonance.

With a further increase in $Z_I$ the relative importance
of the EA channel continues to decrease and
for electron loss from U$^{90+}$ the direct channel
becomes of equal importance even in the vicinity of
the resonance strongly dominating the total emission.

By comparing the results for electron loss from the different ions
in Fig. \ref{fig12} we observe that the shape of the emission pattern
in collisions with atomic particles is very sensitive to the magnitude
of $Z_I$. For instance, comparing electron loss from Ca$^{18+}$
and Xe$^{52+}$ in the vicinity of the resonance corresponding to the
transition $(1s \, 2s)_{J=0} \rightarrow (2s \, 2p_{3/2})_{J=1}$,
where this process in both cases is dominated by the EA channel,
we see that the corresponding angular distributions differ qualitatively.
The reasons for this are the same as was already discussed
in the first subsection of this section: i) the momentum transfer
$q$ to the HCI increases with $Z_I$ faster than the typical
orbiting momenta of the electrons in the HCI, and
ii) the longitudinal component $q_{min}$ of ${\bf q}$ grows
with $Z_I$ somewhat faster that its transverse part $q_{\perp}$.

\subsection{ Spectra of electrons emitted from a moving HCI }

Up to now we discussed the electron emission spectra
by considering them in the rest frame of the HCI.
Such a theoretical discussion would be directly relevant to a possible
experimental exploration of electron loss if the ion rests
in the laboratory frame. The latter can be realized in EBITs
where HCIs are produced by beams of energetic electrons which
can also be utilized as projectiles to induce electron
loss from (already prepared) helium-like highly charged ions.
By combining EBIT with a source of X-rays
(e.g. with a free-electron laser) electron loss
from helium-like ions by photo absorption
could also be experimentally investigated.

On the other hand, a situation in which electron loss from HCIs
is caused by fast collisions with atoms or with (effectively bare)
atomic nuclei can be experimentally realized in accelerators of
heavy ions when the beam of fast HCIs penetrates a gas target
resting in the laboratory frame. Therefore, in
Fig. \ref{fig13} we display results for electron emission from
a moving HCI which are re-calculated to the laboratory frame.
These electrons are emitted from Ca$^{18+}$(1s$^2$) ions which
move in the laboratory frame with an energy of $93$ MeV/u
colliding with a nitrogen target.

In this figure we focus on very small emission angles ($\vartheta \leq 2^\circ$) which are relevant to the properties of the electron spectrometer currently functioning at the GSI (Darmstadt, Germany). Moreover, for these angles we consider only the range of electron emission energies for which autoionizing $(2s\, 2s)_{J=0}$ and $(2s\, 2p_{3/2} )_{J=1}$ states
(which are most visible in the loss process) as well as
the $(2s\, 2p_{1/2})_{J=1}$ state are involved. One can conclude
from the figure that the signatures of autoionization in
electron loss are clearly seen also in the frame
where the HCI moves.

\subsection{ Virtual photon versus real photon }

The relationship between the processes of ionization (electron loss)
by photo absorption and by the impact of a charged particle
is of fundamental importance since its consideration offers
a deeper insight into subtle details of the
response of atoms, ions and molecules
to the action of electromagnetic fields.
As is well known, these two types of ionization (electron loss)
in general qualitatively differ from each other
(this is very clearly seen, for instance, in Figs.
\ref{fig4}--\ref{fig11}).

In our consideration it is assumed that only one photon
is involved in the process of photo electron loss and also that electron loss by charged particle impact proceeds via just a single interaction between the colliding particles. In such a case
any difference between these two electron loss processes
should be fully attributed to the difference between
the properties (momentum and polarization) of a real and virtual photon.

In the case of ionization of light atomic systems,
in which the electron motion is nonrelativistic,
the differences between these properties were discussed
in detail in \cite{voitkiv2001lkn}. There, in particular, it was shown that a virtual photon representing the field of a charged particle
becomes almost real if the following conditions are fulfilled
\begin{eqnarray}
\frac{q_{min}}{ \gamma^2 } \ll q_{\bot} \ll q_{min} \,.
\end{eqnarray}
It can easily be shown that the same holds true also for electron loss from tightly bound systems, like HCIs, where the electron motion
is already relativistic.

The correspondence between electron loss
by photo absorption and by the impact
of a charged particle (proton, electron)
is illustrated in Figs. \ref{fig14} and \ref{fig15} where
the impact energy of the incident proton ranges from a modest relativistic value of $ \approx 94$ MeV ($\gamma = 1.1$) till extreme relativistic energies 
(and the corresponding energies of an equivelocity electron except the lowest one which cannot be treated in the semiclassical approximation).

In Fig. \ref{fig14} these two processes are compared on the basic level
by showing results for the fully differential cross sections.
In Fig. \ref{fig15} results for the energy-angular distribution of
the emitted electrons are displayed. It is seen that
the shapes of the emission pattern in electron loss
by photo absorption and by the charged particle impact
can be practically identical provided the energy of
the incident particle is high enough. 

\section{ Summary }

We have theoretically explored electron loss from helium-like highly charged ions via photo absorption and in collisions with
atomic particles. The main focus of our study was on the role of doubly excited (autoionizing) states of the ion in these processes.
Two initial states of the ion were considered:
the ground $(1s^2)$ state and the metastable
$(1s \, 2s)_{J=0}$ state.

Our description of two-electron states of
a free highly charged ion was based on the line-profile
approach of QED. The interaction of the electrons of
the ion with the incident photon field (with the incident atomic particle)
was treated within the first order of perturbation theory.

Electron loss in general proceeds via two different pathways
which can (partially) interfere. In one of them an electron
of the ion undergoes a direct transition from the initial state
to the continuum. For highly charged ion this -- direct -- channel is essentially a single-electron transition. In the other -- the EA channel -- the ion is excited into an autoionizing state
which then Auger decays.

In electron loss from the ground state the EA channel must
involve in its first step simultaneous transition of both
ionic electrons. This makes this channel highly correlated
because both its excitation and decay parts crucially depend
on the electron-electron interaction. Since the role of
the latter decreases with $Z_I$,
the EA channel in electron loss from the ground state
of highly charged ions in collisions with atomic particles
does not have a noticeable impact on the total loss and
can substantially influence only the emission pattern
in the vicinity of the corresponding resonances.
Moreover, even this influence becomes already quite weak
beginning with $Z_I \approx 35 $ -- $40$.

Electron loss from the ground state in collisions with photons
and atomic particle was explored by considering Ca$^{18+}$
and Zn$^{28+}$ ions. It was, in particular, found that
the shape of the emission pattern
in general strongly depends on whether the process
is caused by photo absorption or by atomic particle impact.
Moreover, in the latter case the shape also very substantially
depends on the atomic number of the HCI.

Autoionizing states with the basic configuration involving
a $2s$-electron can be reached from the metastable
$(1s \, 2s)_{J=0}$ state even without taking
into account the electron-electron interaction.
Therefore, in electron loss
from the $(1s \, 2s)_{J=0}$ state, autoionization plays
a much more prominent role.
In particular, for relatively light and intermediately heavy
helium-like HCIs, where the total width of an autoionizing
state is not yet strongly dominated
by spontaneous radiative decay,
the EA channel has not only an enormous impact
on the emission spectra in the vicinity of the resonances
but also very substantially influences the magnitude of
the total loss cross section. However,
the EA channel becomes much less efficient for
very heavy HCIs because in such ions
the Auger decay is much weaker than
the spontaneous radiative decay.

Electron loss from the metastable $(1s \, 2s)_{J=0}$ state was studied
by performing calculations for Ca$^{18+}$, Xe$^{52+}$ and U$^{90+}$ ions.
Here it was again found that the shape of the emission spectra
in cases of photo absorption and atomic particle impact
in general qualitatively differ and that the shape
substantially (or even qualitatively) changes
when the atomic number of the HCI is varied.

We also compared electron loss from HCIs by photo absorption
and due to the impact of an extreme relativistic charged particle.
Our calculations confirmed the fundamental similarity
of these processes in case when photo absorption is allowed by
selection rules.

\section*{ Acknowledgments }

We are grateful to C. M\"uller for useful discussions.

This work is supported by the German-Russian Interdisciplinary Science Center (G-RISC) funded by the German Federal Foreign Office via the German Academic Exchange Service (DAAD).
K.N.L. acknowledges the support from RFBR Grant 16-32-00620.


\begin{thebibliography}{54}
\expandafter\ifx\csname natexlab\endcsname\relax\def\natexlab#1{#1}\fi
\expandafter\ifx\csname bibnamefont\endcsname\relax
  \def\bibnamefont#1{#1}\fi
\expandafter\ifx\csname bibfnamefont\endcsname\relax
  \def\bibfnamefont#1{#1}\fi
\expandafter\ifx\csname citenamefont\endcsname\relax
  \def\citenamefont#1{#1}\fi
\expandafter\ifx\csname url\endcsname\relax
  \def\url#1{\texttt{#1}}\fi
\expandafter\ifx\csname urlprefix\endcsname\relax\def\urlprefix{URL }\fi
\providecommand{\bibinfo}[2]{#2}
\providecommand{\eprint}[2][]{\url{#2}}

\bibitem[{\citenamefont{Meyer et~al.}(1997)\citenamefont{Meyer, Greene, and
  Esry}}]{meyer1997lkn}
\bibinfo{author}{\bibfnamefont{K.~W.} \bibnamefont{Meyer}},
  \bibinfo{author}{\bibfnamefont{C.~H.} \bibnamefont{Greene}},
  \bibnamefont{and} \bibinfo{author}{\bibfnamefont{B.~D.} \bibnamefont{Esry}},
  \bibinfo{journal}{Phys. Rev. Lett.} \textbf{\bibinfo{volume}{78}},
  \bibinfo{pages}{4902} (\bibinfo{year}{1997}).

\bibitem[{\citenamefont{Carlson}(1967)}]{carlson1967lkn}
\bibinfo{author}{\bibfnamefont{T.~A.} \bibnamefont{Carlson}},
  \bibinfo{journal}{Phys. Rev.} \textbf{\bibinfo{volume}{156}},
  \bibinfo{pages}{142} (\bibinfo{year}{1967}).

\bibitem[{\citenamefont{Kheifets and Bray}(1998)}]{kheifets1998lkn}
\bibinfo{author}{\bibfnamefont{A.~S.} \bibnamefont{Kheifets}} \bibnamefont{and}
  \bibinfo{author}{\bibfnamefont{I.}~\bibnamefont{Bray}},
  \bibinfo{journal}{Phys. Rev. A} \textbf{\bibinfo{volume}{58}},
  \bibinfo{pages}{4501} (\bibinfo{year}{1998}).

\bibitem[{\citenamefont{{D\"orner} et~al.}(1996)\citenamefont{{D\"orner}, Vogt,
  Mergel, Khemliche, Kravis, and et~al.}}]{dorner1996lkn}
\bibinfo{author}{\bibfnamefont{R.}~\bibnamefont{{D\"orner}}},
  \bibinfo{author}{\bibfnamefont{T.}~\bibnamefont{Vogt}},
  \bibinfo{author}{\bibfnamefont{V.}~\bibnamefont{Mergel}},
  \bibinfo{author}{\bibfnamefont{H.}~\bibnamefont{Khemliche}},
  \bibinfo{author}{\bibfnamefont{S.}~\bibnamefont{Kravis}}, \bibnamefont{and}
  \bibinfo{author}{\bibnamefont{et~al.}}, \bibinfo{journal}{Phys. Rev. Lett.}
  \textbf{\bibinfo{volume}{76}}, \bibinfo{pages}{2654} (\bibinfo{year}{1996}).

\bibitem[{\citenamefont{Wang et~al.}(2006)\citenamefont{Wang, Hsiao, and
  Huang}}]{wang2006lkn}
\bibinfo{author}{\bibfnamefont{L.-R.} \bibnamefont{Wang}},
  \bibinfo{author}{\bibfnamefont{J.-T.} \bibnamefont{Hsiao}}, \bibnamefont{and}
  \bibinfo{author}{\bibfnamefont{K.-N.} \bibnamefont{Huang}},
  \bibinfo{journal}{J. Phys. B: At. Mol. Opt. Phys.}
  \textbf{\bibinfo{volume}{39}}, \bibinfo{pages}{L217} (\bibinfo{year}{2006}).

\bibitem[{\citenamefont{Dubau and Wells}(1973)}]{dubau1973lkn}
\bibinfo{author}{\bibfnamefont{J.}~\bibnamefont{Dubau}} \bibnamefont{and}
  \bibinfo{author}{\bibfnamefont{J.}~\bibnamefont{Wells}}, \bibinfo{journal}{J.
  Phys. B: At. Mol. Opt. Phys.} \textbf{\bibinfo{volume}{6}},
  \bibinfo{pages}{L31} (\bibinfo{year}{1973}).

\bibitem[{\citenamefont{Altick}(1968)}]{altick1968lkn}
\bibinfo{author}{\bibfnamefont{P.~L.} \bibnamefont{Altick}},
  \bibinfo{journal}{Phys. Rev.} \textbf{\bibinfo{volume}{169}},
  \bibinfo{pages}{21} (\bibinfo{year}{1968}).

\bibitem[{\citenamefont{Chi et~al.}(1991)\citenamefont{Chi, Huang, and
  Cheng}}]{chi1991lkn}
\bibinfo{author}{\bibfnamefont{H.-C.} \bibnamefont{Chi}},
  \bibinfo{author}{\bibfnamefont{K.-N.} \bibnamefont{Huang}}, \bibnamefont{and}
  \bibinfo{author}{\bibfnamefont{K.~T.} \bibnamefont{Cheng}},
  \bibinfo{journal}{Phys. Rev. A} \textbf{\bibinfo{volume}{43}},
  \bibinfo{pages}{2542} (\bibinfo{year}{1991}).

\bibitem[{\citenamefont{Hsiao et~al.}(2008)\citenamefont{Hsiao, Wang, Sun, Lin,
  Lu, and Huang}}]{hsiao2008lkn}
\bibinfo{author}{\bibfnamefont{J.-T.} \bibnamefont{Hsiao}},
  \bibinfo{author}{\bibfnamefont{L.-R.} \bibnamefont{Wang}},
  \bibinfo{author}{\bibfnamefont{H.-L.} \bibnamefont{Sun}},
  \bibinfo{author}{\bibfnamefont{S.-F.} \bibnamefont{Lin}},
  \bibinfo{author}{\bibfnamefont{C.-L.} \bibnamefont{Lu}}, \bibnamefont{and}
  \bibinfo{author}{\bibfnamefont{K.-N.} \bibnamefont{Huang}},
  \bibinfo{journal}{Phys. Rev. A} \textbf{\bibinfo{volume}{78}},
  \bibinfo{pages}{013411} (\bibinfo{year}{2008}).

\bibitem[{\citenamefont{Wang et~al.}(2014)\citenamefont{Wang, Sheinerman, and
  Robicheaux}}]{wang2014lkn}
\bibinfo{author}{\bibfnamefont{Q.}~\bibnamefont{Wang}},
  \bibinfo{author}{\bibfnamefont{S.}~\bibnamefont{Sheinerman}},
  \bibnamefont{and}
  \bibinfo{author}{\bibfnamefont{F.}~\bibnamefont{Robicheaux}},
  \bibinfo{journal}{J. Phys. B: At. Mol. Opt. Phys.}
  \textbf{\bibinfo{volume}{47}}, \bibinfo{pages}{215003}
  (\bibinfo{year}{2014}).

\bibitem[{\citenamefont{Codling et~al.}(1967)\citenamefont{Codling, Madden, and
  Ederer}}]{codling1967lkn}
\bibinfo{author}{\bibfnamefont{K.}~\bibnamefont{Codling}},
  \bibinfo{author}{\bibfnamefont{R.~P.} \bibnamefont{Madden}},
  \bibnamefont{and} \bibinfo{author}{\bibfnamefont{D.~L.}
  \bibnamefont{Ederer}}, \bibinfo{journal}{Phys. Rev.}
  \textbf{\bibinfo{volume}{155}}, \bibinfo{pages}{26} (\bibinfo{year}{1967}).

\bibitem[{\citenamefont{Wills et~al.}(1998)\citenamefont{Wills, Gorczyca,
  Berrah, Langer, Felfli, Kukk, Bozek, Nayandin, and Alshehri}}]{wills1998lkn}
\bibinfo{author}{\bibfnamefont{A.~A.} \bibnamefont{Wills}},
  \bibinfo{author}{\bibfnamefont{T.}~\bibnamefont{Gorczyca}},
  \bibinfo{author}{\bibfnamefont{N.}~\bibnamefont{Berrah}},
  \bibinfo{author}{\bibfnamefont{B.}~\bibnamefont{Langer}},
  \bibinfo{author}{\bibfnamefont{Z.}~\bibnamefont{Felfli}},
  \bibinfo{author}{\bibfnamefont{E.}~\bibnamefont{Kukk}},
  \bibinfo{author}{\bibfnamefont{J.~D.} \bibnamefont{Bozek}},
  \bibinfo{author}{\bibfnamefont{O.}~\bibnamefont{Nayandin}}, \bibnamefont{and}
  \bibinfo{author}{\bibfnamefont{M.}~\bibnamefont{Alshehri}},
  \bibinfo{journal}{Phys. Rev. Lett.} \textbf{\bibinfo{volume}{80}},
  \bibinfo{pages}{5085} (\bibinfo{year}{1998}).

\bibitem[{\citenamefont{Landers et~al.}(2009)\citenamefont{Landers, Robicheaux,
  Jahnke, {Sch\"offler}, Osipov, Titze, Lee, Adaniya, and
  et~al.}}]{landers2009lkn}
\bibinfo{author}{\bibfnamefont{A.~L.} \bibnamefont{Landers}},
  \bibinfo{author}{\bibfnamefont{F.}~\bibnamefont{Robicheaux}},
  \bibinfo{author}{\bibfnamefont{T.}~\bibnamefont{Jahnke}},
  \bibinfo{author}{\bibfnamefont{M.}~\bibnamefont{{Sch\"offler}}},
  \bibinfo{author}{\bibfnamefont{T.}~\bibnamefont{Osipov}},
  \bibinfo{author}{\bibfnamefont{J.}~\bibnamefont{Titze}},
  \bibinfo{author}{\bibfnamefont{S.}~\bibnamefont{Lee}},
  \bibinfo{author}{\bibfnamefont{H.}~\bibnamefont{Adaniya}}, \bibnamefont{and}
  \bibinfo{author}{\bibnamefont{et~al.}}, \bibinfo{journal}{Phys. Rev. Lett.}
  \textbf{\bibinfo{volume}{102}}, \bibinfo{pages}{223001}
  (\bibinfo{year}{2009}).

\bibitem[{\citenamefont{Lindsay et~al.}(1992)\citenamefont{Lindsay, Cai,
  Schinn, Dai, and Gallagher}}]{lindsay1992lkn}
\bibinfo{author}{\bibfnamefont{M.~D.} \bibnamefont{Lindsay}},
  \bibinfo{author}{\bibfnamefont{L.-T.} \bibnamefont{Cai}},
  \bibinfo{author}{\bibfnamefont{G.~W.} \bibnamefont{Schinn}},
  \bibinfo{author}{\bibfnamefont{C.-J.} \bibnamefont{Dai}}, \bibnamefont{and}
  \bibinfo{author}{\bibfnamefont{T.~F.} \bibnamefont{Gallagher}},
  \bibinfo{journal}{Phys. Rev. A} \textbf{\bibinfo{volume}{45}},
  \bibinfo{pages}{231} (\bibinfo{year}{1992}).

\bibitem[{\citenamefont{Chi and Huang}(1994)}]{chi1994lkn}
\bibinfo{author}{\bibfnamefont{H.-C.} \bibnamefont{Chi}} \bibnamefont{and}
  \bibinfo{author}{\bibfnamefont{K.-N.} \bibnamefont{Huang}},
  \bibinfo{journal}{Phys. Rev. A} \textbf{\bibinfo{volume}{50}},
  \bibinfo{pages}{392} (\bibinfo{year}{1994}).

\bibitem[{\citenamefont{Deshmukh and Johnson}(1983)}]{deshmukh1983lkn}
\bibinfo{author}{\bibfnamefont{P.~C.} \bibnamefont{Deshmukh}} \bibnamefont{and}
  \bibinfo{author}{\bibfnamefont{W.~R.} \bibnamefont{Johnson}},
  \bibinfo{journal}{Phys. Rev. A} \textbf{\bibinfo{volume}{27}},
  \bibinfo{pages}{326} (\bibinfo{year}{1983}).

\bibitem[{\citenamefont{Wang et~al.}(1999)\citenamefont{Wang, Chi, and
  Huang}}]{wang1999lkn}
\bibinfo{author}{\bibfnamefont{L.-R.} \bibnamefont{Wang}},
  \bibinfo{author}{\bibfnamefont{H.-C.} \bibnamefont{Chi}}, \bibnamefont{and}
  \bibinfo{author}{\bibfnamefont{K.-N.} \bibnamefont{Huang}},
  \bibinfo{journal}{Phys. Rev. Lett.} \textbf{\bibinfo{volume}{83}},
  \bibinfo{pages}{702} (\bibinfo{year}{1999}).

\bibitem[{\citenamefont{Bahl et~al.}(1979)\citenamefont{Bahl, Watson, and
  Irgollic}}]{bahl1979lkn}
\bibinfo{author}{\bibfnamefont{M.~K.} \bibnamefont{Bahl}},
  \bibinfo{author}{\bibfnamefont{R.~L.} \bibnamefont{Watson}},
  \bibnamefont{and} \bibinfo{author}{\bibfnamefont{K.~J.}
  \bibnamefont{Irgollic}}, \bibinfo{journal}{Phys. Rev. Lett.}
  \textbf{\bibinfo{volume}{42}}, \bibinfo{pages}{165} (\bibinfo{year}{1979}).

\bibitem[{\citenamefont{Kuchiev and Sheinerman}(1985)}]{kuchiev1985lkn}
\bibinfo{author}{\bibfnamefont{M.~Y.} \bibnamefont{Kuchiev}} \bibnamefont{and}
  \bibinfo{author}{\bibfnamefont{S.~A.} \bibnamefont{Sheinerman}},
  \bibinfo{journal}{J. Phys. B: At. Mol. Opt. Phys.}
  \textbf{\bibinfo{volume}{18}}, \bibinfo{pages}{L551} (\bibinfo{year}{1985}).

\bibitem[{\citenamefont{Madden et~al.}(1969)\citenamefont{Madden, Ederer, and
  Codling}}]{madden1969lkn}
\bibinfo{author}{\bibfnamefont{R.~P.} \bibnamefont{Madden}},
  \bibinfo{author}{\bibfnamefont{D.~L.} \bibnamefont{Ederer}},
  \bibnamefont{and} \bibinfo{author}{\bibfnamefont{K.}~\bibnamefont{Codling}},
  \bibinfo{journal}{Phys. Rev.} \textbf{\bibinfo{volume}{177}},
  \bibinfo{pages}{136} (\bibinfo{year}{1969}).

\bibitem[{\citenamefont{Guillemin et~al.}(2012)\citenamefont{Guillemin,
  Sheinerman, Bomme, Journel, Marin, Marchenko, Kushawaha, Trcera,
  Piancastelli, and Simon}}]{guillemin2012lkn}
\bibinfo{author}{\bibfnamefont{R.}~\bibnamefont{Guillemin}},
  \bibinfo{author}{\bibfnamefont{S.}~\bibnamefont{Sheinerman}},
  \bibinfo{author}{\bibfnamefont{C.}~\bibnamefont{Bomme}},
  \bibinfo{author}{\bibfnamefont{L.}~\bibnamefont{Journel}},
  \bibinfo{author}{\bibfnamefont{T.}~\bibnamefont{Marin}},
  \bibinfo{author}{\bibfnamefont{T.}~\bibnamefont{Marchenko}},
  \bibinfo{author}{\bibfnamefont{R.~K.} \bibnamefont{Kushawaha}},
  \bibinfo{author}{\bibfnamefont{N.}~\bibnamefont{Trcera}},
  \bibinfo{author}{\bibfnamefont{M.~N.} \bibnamefont{Piancastelli}},
  \bibnamefont{and} \bibinfo{author}{\bibfnamefont{M.}~\bibnamefont{Simon}},
  \bibinfo{journal}{Phys. Rev. Lett.} \textbf{\bibinfo{volume}{109}},
  \bibinfo{pages}{013001} (\bibinfo{year}{2012}).

\bibitem[{\citenamefont{Sheinerman et~al.}(2006)\citenamefont{Sheinerman,
  Lablanquie, Penent, Palaudoux, Eland, Aoto, Hikosaka, and
  Ito}}]{sheinerman2006lkn}
\bibinfo{author}{\bibfnamefont{S.}~\bibnamefont{Sheinerman}},
  \bibinfo{author}{\bibfnamefont{P.}~\bibnamefont{Lablanquie}},
  \bibinfo{author}{\bibfnamefont{F.}~\bibnamefont{Penent}},
  \bibinfo{author}{\bibfnamefont{J.}~\bibnamefont{Palaudoux}},
  \bibinfo{author}{\bibfnamefont{J.~H.~D.} \bibnamefont{Eland}},
  \bibinfo{author}{\bibfnamefont{T.}~\bibnamefont{Aoto}},
  \bibinfo{author}{\bibfnamefont{Y.}~\bibnamefont{Hikosaka}}, \bibnamefont{and}
  \bibinfo{author}{\bibfnamefont{K.}~\bibnamefont{Ito}}, \bibinfo{journal}{J.
  Phys. B: At. Mol. Opt. Phys.} \textbf{\bibinfo{volume}{39}},
  \bibinfo{pages}{1017} (\bibinfo{year}{2006}).

\bibitem[{\citenamefont{Kuchiev and Sheinerman}(1994)}]{kuchiev1994lkn}
\bibinfo{author}{\bibfnamefont{M.~Y.} \bibnamefont{Kuchiev}} \bibnamefont{and}
  \bibinfo{author}{\bibfnamefont{S.~A.} \bibnamefont{Sheinerman}},
  \bibinfo{journal}{J. Phys. B: At. Mol. Opt. Phys.}
  \textbf{\bibinfo{volume}{27}}, \bibinfo{pages}{2943} (\bibinfo{year}{1994}).

\bibitem[{\citenamefont{Niehaus}(1977)}]{niehaus1977lkn}
\bibinfo{author}{\bibfnamefont{A.}~\bibnamefont{Niehaus}}, \bibinfo{journal}{J.
  Phys. B: At. Mol. Opt. Phys.} \textbf{\bibinfo{volume}{10}},
  \bibinfo{pages}{1845} (\bibinfo{year}{1977}).

\bibitem[{\citenamefont{Schmidt et~al.}(1977)\citenamefont{Schmidt, Sandner,
  Mehlhorn, Adam, and Wuilleumier}}]{schmidt1977lkn}
\bibinfo{author}{\bibfnamefont{V.}~\bibnamefont{Schmidt}},
  \bibinfo{author}{\bibfnamefont{N.}~\bibnamefont{Sandner}},
  \bibinfo{author}{\bibfnamefont{W.}~\bibnamefont{Mehlhorn}},
  \bibinfo{author}{\bibfnamefont{M.~Y.} \bibnamefont{Adam}}, \bibnamefont{and}
  \bibinfo{author}{\bibfnamefont{F.}~\bibnamefont{Wuilleumier}},
  \bibinfo{journal}{Phys. Rev. Lett.} \textbf{\bibinfo{volume}{38}},
  \bibinfo{pages}{63} (\bibinfo{year}{1977}).

\bibitem[{\citenamefont{Heugel et~al.}(2016)\citenamefont{Heugel, Fischer,
  Elman, Maiwald, Sondermann, and Leuchs}}]{heugel2016lkn}
\bibinfo{author}{\bibfnamefont{S.}~\bibnamefont{Heugel}},
  \bibinfo{author}{\bibfnamefont{M.}~\bibnamefont{Fischer}},
  \bibinfo{author}{\bibfnamefont{V.}~\bibnamefont{Elman}},
  \bibinfo{author}{\bibfnamefont{R.}~\bibnamefont{Maiwald}},
  \bibinfo{author}{\bibfnamefont{M.}~\bibnamefont{Sondermann}},
  \bibnamefont{and} \bibinfo{author}{\bibfnamefont{G.}~\bibnamefont{Leuchs}},
  \bibinfo{journal}{J. Phys. B: At. Mol. Opt. Phys.}
  \textbf{\bibinfo{volume}{49}}, \bibinfo{pages}{015002}
  (\bibinfo{year}{2016}).

\bibitem[{\citenamefont{Madden and Codling}(1965)}]{madden1965lkn}
\bibinfo{author}{\bibfnamefont{R.~P.} \bibnamefont{Madden}} \bibnamefont{and}
  \bibinfo{author}{\bibfnamefont{K.}~\bibnamefont{Codling}},
  \bibinfo{journal}{Astrophys. J.} \textbf{\bibinfo{volume}{141}},
  \bibinfo{pages}{364} (\bibinfo{year}{1965}).

\bibitem[{\citenamefont{Ormonde et~al.}(1967)\citenamefont{Ormonde, Whitaker,
  and Lipsky}}]{ormonde1967lkn}
\bibinfo{author}{\bibfnamefont{S.}~\bibnamefont{Ormonde}},
  \bibinfo{author}{\bibfnamefont{W.}~\bibnamefont{Whitaker}}, \bibnamefont{and}
  \bibinfo{author}{\bibfnamefont{L.}~\bibnamefont{Lipsky}},
  \bibinfo{journal}{Phys. Rev. Lett.} \textbf{\bibinfo{volume}{19}},
  \bibinfo{pages}{1161} (\bibinfo{year}{1967}).

\bibitem[{\citenamefont{Raeker et~al.}(1994)\citenamefont{Raeker, Bartschat,
  and Reid}}]{raeker1994lkn}
\bibinfo{author}{\bibfnamefont{A.}~\bibnamefont{Raeker}},
  \bibinfo{author}{\bibfnamefont{K.}~\bibnamefont{Bartschat}},
  \bibnamefont{and} \bibinfo{author}{\bibfnamefont{R.~H.~G.}
  \bibnamefont{Reid}}, \bibinfo{journal}{J. Phys. B: At. Mol. Opt. Phys.}
  \textbf{\bibinfo{volume}{27}}, \bibinfo{pages}{3129} (\bibinfo{year}{1994}).

\bibitem[{\citenamefont{Falk et~al.}(1981)\citenamefont{Falk, Dunn, Griffin,
  Bottcher, Gregory, Crandall, and Pindzola}}]{falk1981lkn}
\bibinfo{author}{\bibfnamefont{R.~A.} \bibnamefont{Falk}},
  \bibinfo{author}{\bibfnamefont{G.~H.} \bibnamefont{Dunn}},
  \bibinfo{author}{\bibfnamefont{D.~C.} \bibnamefont{Griffin}},
  \bibinfo{author}{\bibfnamefont{C.}~\bibnamefont{Bottcher}},
  \bibinfo{author}{\bibfnamefont{D.~C.} \bibnamefont{Gregory}},
  \bibinfo{author}{\bibfnamefont{D.~H.} \bibnamefont{Crandall}},
  \bibnamefont{and} \bibinfo{author}{\bibfnamefont{M.~S.}
  \bibnamefont{Pindzola}}, \bibinfo{journal}{Phys. Rev. Lett.}
  \textbf{\bibinfo{volume}{47}}, \bibinfo{pages}{494} (\bibinfo{year}{1981}).

\bibitem[{\citenamefont{Peart et~al.}(1973)\citenamefont{Peart, Stevenson, and
  Dolder}}]{peart1973lkn}
\bibinfo{author}{\bibfnamefont{B.}~\bibnamefont{Peart}},
  \bibinfo{author}{\bibfnamefont{J.~G.} \bibnamefont{Stevenson}},
  \bibnamefont{and} \bibinfo{author}{\bibfnamefont{K.~T.}
  \bibnamefont{Dolder}}, \bibinfo{journal}{J. Phys B}
  \textbf{\bibinfo{volume}{6}}, \bibinfo{pages}{146} (\bibinfo{year}{1973}).

\bibitem[{\citenamefont{Dolder and Peart}(1976)}]{dolder1976lkn}
\bibinfo{author}{\bibfnamefont{K.~T.} \bibnamefont{Dolder}} \bibnamefont{and}
  \bibinfo{author}{\bibfnamefont{B.}~\bibnamefont{Peart}},
  \bibinfo{journal}{Rep. Prog. Phys.} \textbf{\bibinfo{volume}{39}},
  \bibinfo{pages}{693} (\bibinfo{year}{1976}).

\bibitem[{\citenamefont{R.~K.~Feeney and Elford}(1972)}]{feeney1972lkn}
\bibinfo{author}{\bibinfo{author}{\bibfnamefont{R.~K.} \bibnamefont{Feeney}}},
\bibinfo{author}{\bibfnamefont{J.~W.}\bibnamefont{Hooper}}
  \bibnamefont{and} \bibinfo{author}{\bibfnamefont{M.~T.}
  \bibnamefont{Elford}}, \bibinfo{journal}{Phys. Rev. A.}
  \textbf{\bibinfo{volume}{6}}, \bibinfo{pages}{1469} (\bibinfo{year}{1972}).

\bibitem[{\citenamefont{Crandall et~al.}(1979)\citenamefont{Crandall, Phaneuf,
  Hasselquist, and Gregory}}]{crandall1979lkn}
\bibinfo{author}{\bibfnamefont{D.~H.} \bibnamefont{Crandall}},
  \bibinfo{author}{\bibfnamefont{R.~A.} \bibnamefont{Phaneuf}},
  \bibinfo{author}{\bibfnamefont{B.~E.} \bibnamefont{Hasselquist}},
  \bibnamefont{and} \bibinfo{author}{\bibfnamefont{D.~C.}
  \bibnamefont{Gregory}}, \bibinfo{journal}{J. Phys. B}
  \textbf{\bibinfo{volume}{12}}, \bibinfo{pages}{L249} (\bibinfo{year}{1979}).

\bibitem[{\citenamefont{{M\"uller} et~al.}(1989)\citenamefont{{M\"uller},
  Hofmann, Weissbecker, Stenke, Tinschert, Wagner, and
  Salzborn}}]{muller1989lkn}
\bibinfo{author}{\bibfnamefont{A.}~\bibnamefont{{M\"uller}}},
  \bibinfo{author}{\bibfnamefont{G.}~\bibnamefont{Hofmann}},
  \bibinfo{author}{\bibfnamefont{B.}~\bibnamefont{Weissbecker}},
  \bibinfo{author}{\bibfnamefont{M.}~\bibnamefont{Stenke}},
  \bibinfo{author}{\bibfnamefont{K.}~\bibnamefont{Tinschert}},
  \bibinfo{author}{\bibfnamefont{M.}~\bibnamefont{Wagner}}, \bibnamefont{and}
  \bibinfo{author}{\bibfnamefont{E.}~\bibnamefont{Salzborn}},
  \bibinfo{journal}{Phys. Rev. Lett.} \textbf{\bibinfo{volume}{63}},
  \bibinfo{pages}{758} (\bibinfo{year}{1989}).

\bibitem[{\citenamefont{Henry}(1979)}]{henry1979lkn}
\bibinfo{author}{\bibfnamefont{H.~J.~W.} \bibnamefont{Henry}},
  \bibinfo{journal}{J. Phys. B} \textbf{\bibinfo{volume}{12}},
  \bibinfo{pages}{L309} (\bibinfo{year}{1979}).

\bibitem[{\citenamefont{Tayal and Henry}(1991)}]{tayal1991lkn}
\bibinfo{author}{\bibfnamefont{S.~S.} \bibnamefont{Tayal}} \bibnamefont{and}
  \bibinfo{author}{\bibfnamefont{R.~J.~W.} \bibnamefont{Henry}},
  \bibinfo{journal}{Phys. Rev. A.} \textbf{\bibinfo{volume}{44}},
  \bibinfo{pages}{2955} (\bibinfo{year}{1991}).

\bibitem[{\citenamefont{Pindzola et~al.}(1982)\citenamefont{Pindzola,
  C.Griffin, and Bottcher}}]{pindzola1982lkn}
\bibinfo{author}{\bibfnamefont{M.~S.} \bibnamefont{Pindzola}},
  \bibinfo{author}{\bibfnamefont{D.}~\bibnamefont{C.Griffin}},
  \bibnamefont{and} \bibinfo{author}{\bibfnamefont{C.}~\bibnamefont{Bottcher}},
  \bibinfo{journal}{Phys. Rev. A.} \textbf{\bibinfo{volume}{25}},
  \bibinfo{pages}{211} (\bibinfo{year}{1982}).

\bibitem[{\citenamefont{Ballance et~al.}(2011)\citenamefont{Ballance, Loch,
  Ludlow, Abdel-Naby, and Pindzola}}]{ballance2011lkn}
\bibinfo{author}{\bibfnamefont{C.~P.} \bibnamefont{Ballance}},
  \bibinfo{author}{\bibfnamefont{S.~D.} \bibnamefont{Loch}},
  \bibinfo{author}{\bibfnamefont{J.~A.} \bibnamefont{Ludlow}},
  \bibinfo{author}{\bibfnamefont{S.~A.} \bibnamefont{Abdel-Naby}},
  \bibnamefont{and} \bibinfo{author}{\bibfnamefont{M.~S.}
  \bibnamefont{Pindzola}}, \bibinfo{journal}{Phys. Rev. A.}
  \textbf{\bibinfo{volume}{84}}, \bibinfo{pages}{062713}
  (\bibinfo{year}{2011}).

\bibitem[{\citenamefont{Griffin et~al.}(1984)\citenamefont{Griffin, Bottcher,
  and Pindzola}}]{griffin1984lkn}
\bibinfo{author}{\bibfnamefont{D.~C.} \bibnamefont{Griffin}},
  \bibinfo{author}{\bibfnamefont{C.}~\bibnamefont{Bottcher}}, \bibnamefont{and}
  \bibinfo{author}{\bibfnamefont{M.~S.} \bibnamefont{Pindzola}},
  \bibinfo{journal}{Phys. Rev. A} \textbf{\bibinfo{volume}{29}},
  \bibinfo{pages}{1729} (\bibinfo{year}{1984}).

\bibitem[{\citenamefont{Jr. et~al.}(2011)\citenamefont{Jr., Brandau, Jacobi, Schippers, and Muller}}]{borovik2011lkn}
\bibinfo{author}{\bibfnamefont{A.} \bibnamefont{Borovik} \bibfnamefont{Jr.}},
  \bibinfo{author}{\bibfnamefont{C.}~\bibnamefont{Brandau}},
  \bibinfo{author}{\bibfnamefont{J.}~\bibnamefont{Jacobi}},
  \bibinfo{author}{\bibfnamefont{S.}~\bibnamefont{Schippers}},
  \bibnamefont{and} \bibinfo{author}{\bibfnamefont{A.}~
  \bibnamefont{{M\"uller}}},
  \bibinfo{journal}{J. Phys. B: At. Mol. Opt. Phys.}
  \textbf{\bibinfo{volume}{44}}, \bibinfo{pages}{205205}
  (\bibinfo{year}{2011}).

\bibitem[{\citenamefont{Chen and Reed}(1993)}]{chen1993lkn}
\bibinfo{author}{\bibfnamefont{M.~H.} \bibnamefont{Chen}} \bibnamefont{and}
  \bibinfo{author}{\bibfnamefont{K.~J.} \bibnamefont{Reed}},
  \bibinfo{journal}{Phys. Rev. A.} \textbf{\bibinfo{volume}{48}},
  \bibinfo{pages}{1129} (\bibinfo{year}{1993}).

\bibitem[{\citenamefont{Borovik et~al.}(2013)\citenamefont{Borovik, Gharaibeh,
  Hillenbrand, and {M\"uller}}}]{borovik2013kn}
\bibinfo{author}{\bibfnamefont{A.}~\bibnamefont{Borovik}~\bibfnamefont{Jr.}},
  \bibinfo{author}{\bibfnamefont{M.~F.} \bibnamefont{Gharaibeh}},
  \bibinfo{author}{\bibfnamefont{P.~M.} \bibnamefont{Hillenbrand}},
  \bibnamefont{and} \bibinfo{author}{\bibfnamefont{S.~S.~A.}
  \bibnamefont{{M\"uller}}}, \bibinfo{journal}{J. Phys. B: At. Mol. Opt. Phys.}
  \textbf{\bibinfo{volume}{46}}, \bibinfo{pages}{175201}
  (\bibinfo{year}{2013}).

\bibitem[{\citenamefont{Fritzsche et~al.}(2012)\citenamefont{Fritzsche,
  Surzhykov, Gumberidze, and Stohlker}}]{fritzsche2012lkn}
\bibinfo{author}{\bibfnamefont{S.}~\bibnamefont{Fritzsche}},
  \bibinfo{author}{\bibfnamefont{A.}~\bibnamefont{Surzhykov}},
  \bibinfo{author}{\bibfnamefont{A.}~\bibnamefont{Gumberidze}},
  \bibnamefont{and} \bibinfo{author}{\bibfnamefont{T.}~\bibnamefont{Stohlker}},
  \bibinfo{journal}{New J. of Phys.} \textbf{\bibinfo{volume}{14}},
  \bibinfo{pages}{083018} (\bibinfo{year}{2012}).

\bibitem[{\citenamefont{{M\"uller}}(2008)}]{muller2008lkn_book}
\bibinfo{author}{\bibfnamefont{A.}~\bibnamefont{{M\"uller}}},
  \emph{\bibinfo{title}{{Electronion collisions: fundamental processes in
  the focus of applied research}}} (\bibinfo{publisher}{{Elsevier}},
  \bibinfo{year}{2008}).

\bibitem[{\citenamefont{Andreev et~al.}(2009)\citenamefont{Andreev, Labzowsky,
  and Prigorovsky}}]{andreev2009lkn}
\bibinfo{author}{\bibfnamefont{O.~Y.} \bibnamefont{Andreev}},
  \bibinfo{author}{\bibfnamefont{L.~N.} \bibnamefont{Labzowsky}},
  \bibnamefont{and} \bibinfo{author}{\bibfnamefont{A.~V.}
  \bibnamefont{Prigorovsky}}, \bibinfo{journal}{Phys. Rev. A}
  \textbf{\bibinfo{volume}{80}}, \bibinfo{pages}{042514}
  (\bibinfo{year}{2009}).

\bibitem[{\citenamefont{Andreev et~al.}(2008)\citenamefont{Andreev, Labzowsky,
  Plunien, and Solovyev}}]{andreev2008lkn}
\bibinfo{author}{\bibfnamefont{O.~Y.} \bibnamefont{Andreev}},
  \bibinfo{author}{\bibfnamefont{L.~N.} \bibnamefont{Labzowsky}},
  \bibinfo{author}{\bibfnamefont{G.}~\bibnamefont{Plunien}}, \bibnamefont{and}
  \bibinfo{author}{\bibfnamefont{D.~A.} \bibnamefont{Solovyev}},
  \bibinfo{journal}{Phys. Rep.} \textbf{\bibinfo{volume}{455}},
  \bibinfo{pages}{135} (\bibinfo{year}{2008}).

\bibitem[{\citenamefont{akhiezer}(1996)}]{akhiezer1965}
\bibinfo{author}{\bibfnamefont{A.~I.} \bibnamefont{Akhiezer}},
\bibinfo{author}{\bibfnamefont{V.~B.} \bibnamefont{Berestetskii}},
  \emph{ \bibinfo{title}{{Quantum Electrodynamics}}} (\bibinfo{publisher}{Wiley Interscience}, \bibinfo{address}{New York},
  \bibinfo{year}{1965}).

\bibitem[{\citenamefont{Labzowsky}(1996)}]{labzowsky96b}
\bibinfo{author}{\bibfnamefont{L.~N.} \bibnamefont{Labzowsky}},
  \emph{\bibinfo{title}{{Teoriya atoma. Kvantovaya elektrodinamika elektronnyh
  obolochek i processy izlucheniya {\rm[}Theory of atoms. Quantum
  electrodynamics of the electron shells and the processes of radiation{\rm]}
  (in Russian)}}} (\bibinfo{publisher}{Nauka}, \bibinfo{address}{Moscow},
  \bibinfo{year}{1996}).

\bibitem[{\citenamefont{{N. Stolterfoht and R. D. DuBois and R.D.Rivarola}}(1997)}]{stolterfoht1997lkn_book}
\bibinfo{author}{\bibfnamefont{N.}~\bibnamefont{Stolterfoht}},
\bibinfo{author}{\bibfnamefont{R.~D.}~\bibnamefont{DuBois}} \bibnamefont{and}
\bibinfo{author}{\bibfnamefont{R.~D.}~\bibnamefont{Rivarola}},
  \emph{ \bibinfo{title}{{Electron Emission in Heavy Ion-Atom Collisions}}} (\bibinfo{publisher}{{Springer}},
  \bibinfo{year}{1997}).

\bibitem[{\citenamefont{{J. H. McGuire}}(1997)}]{mcguire1997lkn_book}
\bibinfo{author}{\bibfnamefont{J.~H.}~\bibnamefont{McGuire}},
  \emph{\bibinfo{title}{{Electron Correlation Dynamics in Atomic Collisions}}} (\bibinfo{publisher}{{Cambridge University Press}},
  \bibinfo{year}{1997}).



\bibitem[{\citenamefont{Voitkiv and Ullrich}(2008)}]{voitkiv2008lkn}
\bibinfo{author}{\bibfnamefont{A.~B.} \bibnamefont{Voitkiv}} \bibnamefont{and}
  \bibinfo{author}{\bibfnamefont{J.}~\bibnamefont{Ullrich}},
  \emph{\bibinfo{title}{{Relativistic Collisions of Structered Atomic
  Particles}}} (\bibinfo{publisher}{{Springer}}, \bibinfo{address}{Berlin,
  Heidelberg}, \bibinfo{year}{2008}).

\bibitem{antiscreening} Since the effect of the atomic electrons
in the inelastic atomic mode is opposite to that in the elastic one,
the inelastic mode is often called antiscreening, see e.g.
\cite{stolterfoht1997lkn_book,mcguire1997lkn_book,voitkiv2008lkn}.


\bibitem[{\citenamefont{Moliere}(1947)}]{moliere1947lkn}
\bibinfo{author}{\bibfnamefont{G.}~\bibnamefont{Moliere}},
  \bibinfo{journal}{Naturforsch} \textbf{\bibinfo{volume}{2A}},
  \bibinfo{pages}{133} (\bibinfo{year}{1947}).

\bibitem[{\citenamefont{Salvat et~al.}(1987)\citenamefont{Salvat, Martinez,
  Mayol, and Parellada}}]{salvat1987lkn}
\bibinfo{author}{\bibfnamefont{F.}~\bibnamefont{Salvat}},
  \bibinfo{author}{\bibfnamefont{J.~D.} \bibnamefont{Martinez}},
  \bibinfo{author}{\bibfnamefont{R.}~\bibnamefont{Mayol}}, \bibnamefont{and}
  \bibinfo{author}{\bibfnamefont{J.}~\bibnamefont{Parellada}},
  \bibinfo{journal}{Phys. Rev. A} \textbf{\bibinfo{volume}{36}},
  \bibinfo{pages}{467} (\bibinfo{year}{1987}).

\bibitem{nitrogen} When considering collisions with neutral nitrogen
we give cross sections per atom such that in case of N$_2$ these cross sections should be multiplied by a factor of $2$.

\bibitem{strongest_resonance} The autoionizing
$(2s \, 2p_{3/2})_{J=1}$ state leads to the strongest resonance
when the collision velocity substantially exceeds
the typical orbiting velocities of the electrons in the HCI.

\bibitem[{\citenamefont{Fano}(1961)}]{fano1961lkn}
\bibinfo{author}{\bibfnamefont{U.}~\bibnamefont{Fano}}, \bibinfo{journal}{Phys.
  Rev.} \textbf{\bibinfo{volume}{124}}, \bibinfo{pages}{1866}
  (\bibinfo{year}{1961}).

\bibitem{footnote} Momenta transfers involved in these collisions
are so large on the scale of nitrogen that the screening effect
of atomic/molecular electrons can be neglected.

\bibitem[{\citenamefont{Voitkiv and Ullrich}(2001)}]{voitkiv2001lkn}
\bibinfo{author}{\bibfnamefont{A.~B.} \bibnamefont{Voitkiv}} \bibnamefont{and}
  \bibinfo{author}{\bibfnamefont{J.}~\bibnamefont{Ullrich}},
  \bibinfo{journal}{J. Phys. B: At. Mol. Opt. Phys.}
  \textbf{\bibinfo{volume}{34}}, \bibinfo{pages}{4513} (\bibinfo{year}{2001}).

\end{thebibliography}

\begin{table}
\caption{ The position of the resonances, $E_{res} = E_a - \varepsilon_{1s} - m_e c^2$ and the total width, $\Gamma$, of the $(2s2p_{3/2})_1$ autoionizing state for electron loss from Ca$^{18+}$, Zn$^{28+}$,
Xe$^{52+}$ and U$^{90+}$. The results in the second column were obtained by neglecting vacuum polarization and self-energy corrections. The fourth column presents the total width. }
\label{table1}
\begin{center}
\begin{tabular}{c|c|c|c}
\hline
&\phantom{12} $E_{res}$ \phantom{12}& \phantom{12} $E_{res}$  \phantom{12} & $\Gamma$ \\

& keV& keV& eV \\

\hline
\ttfamily Ca$^{18+}$  &2.8407 &2.8393 &0.16\\
\hline
\ttfamily Zn$^{28+}$  &6.3664 & 6.3605&0.44\\
\hline
\ttfamily Xe$^{52+}$ &21.152&21.117&3.5\\
\hline
\ttfamily U$^{90+}$ &68.757&68.55 &26.27\\

\end{tabular}
\end{center}
\end{table}

\vspace{5cm}

\unitlength=1.00mm                          
\special{em:linewidth 0.5pt}
\linethickness{0.5pt}
\begin{picture}(50.00,45.00)(00.00,-5.00)
\put(69.0,05.00){\line(2,1){59.0}}
\put(129.0,35.00){\line(-4,-1){6.0}}
\put(128.0,35.00){\line(-3,-4){3.0}}
\put(110.00,15.00){\makebox(0,0)[cb]{direct ionization}}%
\put(110.50,35.00){\line(-2,1){4.00}}
\put(110.50,35.00){\line(-2,-1){4.0}}
\put(90.0,35.00){\line(1,0){20.00}}
\put(100.00,37.00){\makebox(0,0)[cb]{Auger decay}}%
\put(120.0,35.00){\line(1,0){20.0}}
\put(59.0,35.00){\line(1,0){20.00}}
\put(69.00,37.00){\makebox(0,0)[cb]{autoionizing state}}%
\put(46.0,20.00){\line(-1,2){1.80}}
\put(48.0,20.00){\line(-1,2){1.80}}
\put(50.0,20.00){\line(-1,2){1.80}}
\put(52.0,20.00){\line(-1,2){1.80}}
\put(54.0,20.00){\line(-1,2){1.80}}
\put(56.0,20.00){\line(-1,2){1.80}}
\put(58.0,20.00){\line(-1,2){1.80}}
\put(60.0,20.00){\line(-1,2){1.80}}
\put(62.0,20.00){\line(-1,2){1.80}}
\put(64.0,20.00){\line(-1,2){1.80}}
\put(66.0,20.00){\line(-1,2){1.80}}
\put(68.0,20.00){\line(-1,2){1.80}}
\put(70.0,20.00){\line(-1,2){1.80}}
\put(72.0,20.00){\line(-1,2){1.80}}
\put(74.0,20.00){\line(-1,2){1.80}}
\put(76.0,20.00){\line(-1,2){1.80}}
\put(78.0,20.00){\line(-1,2){1.80}}
\put(80.0,20.00){\line(-1,2){1.80}}
\put(82.0,20.00){\line(-1,2){1.80}}
\put(84.0,20.00){\line(-1,2){1.80}}
\put(86.0,20.00){\line(-1,2){1.80}}
\put(88.0,20.00){\line(-1,2){1.80}}
\put(90.0,20.00){\line(-1,2){1.80}}
\put(92.0,20.00){\line(-1,2){1.80}}
\put(94.0,20.00){\line(-1,2){1.80}}
\put(96.0,20.00){\line(-1,2){1.80}}
\put(98.0,20.00){\line(-1,2){1.80}}
\put(100.0,20.00){\line(-1,2){1.80}}
\put(102.0,20.00){\line(-1,2){1.80}}
\put(104.0,20.00){\line(-1,2){1.80}}
\put(106.0,20.00){\line(-1,2){1.80}}
\put(108.0,20.00){\line(-1,2){1.80}}
\put(110.0,20.00){\line(-1,2){1.80}}
\put(112.0,20.00){\line(-1,2){1.80}}
\put(114.0,20.00){\line(-1,2){1.80}}
\put(116.0,20.00){\line(-1,2){1.80}}
\put(118.0,20.00){\line(-1,2){1.80}}
\put(120.0,20.00){\line(-1,2){1.80}}
\put(122.0,20.00){\line(-1,2){1.80}}
\put(124.0,20.00){\line(-1,2){1.80}}
\put(126.0,20.00){\line(-1,2){1.80}}
\put(128.0,20.00){\line(-1,2){1.80}}
\put(130.0,20.00){\line(-1,2){1.80}}
\put(132.0,20.00){\line(-1,2){1.80}}
\put(134.0,20.00){\line(-1,2){1.80}}
\put(136.0,20.00){\line(-1,2){1.80}}
\put(138.0,20.00){\line(-1,2){1.80}}
\put(140.0,20.00){\line(-1,2){1.80}}
\put(142.0,20.00){\line(-1,2){1.80}}
\put(144.0,20.00){\line(-1,2){1.80}}
\put(146.0,20.00){\line(-1,2){1.80}}
\put(148.0,20.00){\line(-1,2){1.80}}
\put(44.0,20.00){\line(1,0){104.00}}
\put(69.0,35.00){\line(-1,-2){1.80}}
\put(69.0,35.00){\line(1,-2){1.80}}
\put(69.0,05.00){\line(0,1){30.00}}
\put(59.0,05.00){\line(1,0){20.00}}
\put(69.00,1.00){\makebox(0,0)[cb]{initial state}}%

\put(64.0,12.50){\line(-2, 1){04.00}}
\put(64.0,12.50){\line(-2,-1){04.00}}

\put(53.50,14.50){\makebox(0,0)[cb]{${\bf{q}}$}}%
\bezier{40}(43.00,12.50)(44.41,13.32)(45.82,12.50)
\bezier{40}(45.82,12.50)(47.23,9.68)(48.64,12.50)
\bezier{40}(48.64,12.50)(50.05,13.32)(51.46,12.50)
\bezier{40}(51.46,12.50)(52.87,9.68)(54.28,12.50)
\bezier{40}(54.28,12.50)(55.69,13.32)(57.00,12.50)
\bezier{40}(57.00,12.50)(58.41,9.68)(59.82,12.50)
\bezier{40}(59.82,12.50)(61.23,13.32)(62.64,12.50)

\end{picture}

\begin{figure}[h]
\begin{minipage}{40pc}
\caption{ Electron loss from HCI by a single interaction
with an external field. }
\label{fig1}
\end{minipage}\hspace{2pc}%
\end{figure}

\vspace{5cm}

\begin{center}
\unitlength=1.0mm                          
\special{em:linewidth 0.5pt}
\linethickness{0.5pt}

\begin{picture}(50.00,45.00)(0.00,-5.00)
\put(04.60,05.00){\line(0,1){30.00}}
\put(05.40,05.00){\line(0,1){30.00}}
\put(05.00,01.00){\makebox(0,0)[cb]{$1s$}}
\put(05.00,39.00){\makebox(0,0)[ct]{$1s$}}
\put(24.60,05.00){\line(0,1){30.00}}
\put(25.40,05.00){\line(0,1){30.00}}
\put(25.00,30.00){\circle*{1.00}}
\put(25.00,01.00){\makebox(0,0)[cb]{$1s$}}
\put(25.00,39.00){\makebox(0,0)[ct]{$e^-$}}
\put(26.00,20.00){\makebox(0,0)[lc]{$$}}
%
%
\bezier{40}(25.00,30.00)(26.41,32.82)(27.82,30.00)
\bezier{40}(27.82,30.00)(29.23,27.18)(30.64,30.00)
\bezier{40}(30.64,30.00)(32.05,32.82)(33.46,30.00)
\bezier{40}(33.46,30.00)(34.87,27.18)(36.28,30.00)
\bezier{40}(36.28,30.00)(37.69,32.82)(39.00,30.00)
%
\put( 38.80,30.30){\line(1, -1){04.00}}
\put( 38.80,30.30){\line(1,1){04.00}}
\put( 38.80,30.30){\line(-1, 1){04.00}}
\put( 38.80,30.30){\line(-1,-1){04.00}}
\put( 30.00,26.00){\makebox(0,0)[lc]{${\bf{q}}$}}
\put(15.00,-02.00){\makebox(0,0)[cb]{$a$}}
\end{picture}

\begin{picture}(50.00,45.00)(0.00,-5.00)
\put(04.60,05.00){\line(0,1){30.00}}
\put(05.40,05.00){\line(0,1){30.00}}
\put(05.00,10.00){\circle*{1.00}}
\put(05.00,01.00){\makebox(0,0)[cb]{$1s$}}
\put(05.00,39.00){\makebox(0,0)[ct]{$1s$}}
\put(24.60,05.00){\line(0,1){30.00}}
\put(25.40,05.00){\line(0,1){30.00}}
\put(25.00,30.00){\circle*{1.00}}
\put(25.00,10.00){\circle*{1.00}}
\put(25.00,01.00){\makebox(0,0)[cb]{$1s$}}
\put(25.00,39.00){\makebox(0,0)[ct]{$e^-$}}
\put(26.00,20.00){\makebox(0,0)[lc]{$$}}
\bezier{40}(05.00,10.00)(06.41,12.82)(07.82,10.00)
\bezier{40}(07.82,10.00)(09.23,07.18)(10.64,10.00)
\bezier{40}(10.64,10.00)(12.05,12.82)(13.46,10.00)
\bezier{40}(13.46,10.00)(14.87,07.18)(16.28,10.00)
\bezier{40}(16.28,10.00)(17.69,12.82)(19.00,10.00)
\bezier{40}(19.00,10.00)(20.41,07.18)(21.82,10.00)
\bezier{40}(21.82,10.00)(23.23,12.82)(24.64,10.00)
\bezier{40}(25.00,30.00)(26.41,32.82)(27.82,30.00)
\bezier{40}(27.82,30.00)(29.23,27.18)(30.64,30.00)
\bezier{40}(30.64,30.00)(32.05,32.82)(33.46,30.00)
\bezier{40}(33.46,30.00)(34.87,27.18)(36.28,30.00)
\bezier{40}(36.28,30.00)(37.69,32.82)(39.00,30.00)
%
\put( 39.00,30.30){\line(1, -1){04.00}}
\put( 39.00,30.30){\line(1,1){04.00}}
\put( 39.00,30.30){\line(-1,1){04.00}}
\put( 39.00,30.30){\line(-1,-1){04.00}}
\put( 30.00,26.00){\makebox(0,0)[lc]{${\bf{q}}$}}
\put(15.00,-02.00){\makebox(0,0)[cb]{$b$}}
\end{picture}
\begin{picture}(50.00,45.00)(0.00,-5.00)
%
%
\put(24.60,05.00){\line(0,1){30.00}}
\put(25.40,05.00){\line(0,1){30.00}}
\put(25.00,30.00){\circle*{1.00}}
\put(25.00,10.00){\circle*{1.00}}
\put(25.00,01.00){\makebox(0,0)[cb]{$1s $}}
\put(25.00,39.00){\makebox(0,0)[ct]{$1s$}}
\put(26.00,20.00){\makebox(0,0)[lc]{$$}}
\put(44.60,05.00){\line(0,1){30.00}}
\put(45.40,05.00){\line(0,1){30.00}}
\put(45.00,30.00){\circle*{1.00}}
\put(45.00,01.00){\makebox(0,0)[cb]{$1s$}}
\put(45.00,39.00){\makebox(0,0)[ct]{$e^-$}}
%
%
\bezier{40}(10.64,10.00)(12.05,12.82)(13.46,10.00)
\bezier{40}(13.46,10.00)(14.87,07.18)(16.28,10.00)
\bezier{40}(16.28,10.00)(17.69,12.82)(19.00,10.00)
\bezier{40}(19.00,10.00)(20.41,07.18)(21.82,10.00)
\bezier{40}(21.82,10.00)(23.23,12.82)(24.64,10.00)
\bezier{40}(25.00,30.00)(26.41,32.82)(27.82,30.00)
\bezier{40}(27.82,30.00)(29.23,27.18)(30.64,30.00)
\bezier{40}(30.64,30.00)(32.05,32.82)(33.46,30.00)
\bezier{40}(33.46,30.00)(34.87,27.18)(36.28,30.00)
\bezier{40}(36.28,30.00)(37.69,32.82)(39.00,30.00)
\bezier{40}(39.00,30.00)(40.41,27.18)(41.82,30.00)
\bezier{40}(41.82,30.00)(43.23,32.82)(44.64,30.00)
\put(11.10, 10.10){\line(-1, -1){04.00}}
\put(11.10, 10.10){\line(-1,1){04.00}}
\put(11.10, 10.10){\line(1, 1){04.00}}
\put(11.10, 10.10){\line(1,-1){04.00}}
\put( 18.00,15.00){\makebox(0,0)[lc]{${\bf{q}}$}}
\put(35.00,-02.00){\makebox(0,0)[cb]{$c$}}
\end{picture}

\begin{picture}(50.00,45.00)(0.00,-5.00)
\put(04.60,05.00){\line(0,1){30.00}}
\put(05.40,05.00){\line(0,1){30.00}}
\put(05.00,10.00){\circle*{1.00}}
\put(05.00,01.00){\makebox(0,0)[cb]{$1s$}}
\put(05.00,39.00){\makebox(0,0)[ct]{$1s$}}
\put(24.60,05.00){\line(0,1){30.00}}
\put(25.40,05.00){\line(0,1){30.00}}
\put(4.9,30.00){\circle*{1.00}}
\put(25,10.00){\circle*{1.00}}
\put(25.00,01.00){\makebox(0,0)[cb]{$1s$}}
\put(25.00,39.00){\makebox(0,0)[ct]{$e^-$}}
\put(26.00,20.00){\makebox(0,0)[lc]{$$}}
\bezier{40}(05.00,10.00)(06.41,12.82)(07.82,10.00)
\bezier{40}(07.82,10.00)(09.23,07.18)(10.64,10.00)
\bezier{40}(10.64,10.00)(12.05,12.82)(13.46,10.00)
\bezier{40}(13.46,10.00)(14.87,07.18)(16.28,10.00)
\bezier{40}(16.28,10.00)(17.69,12.82)(19.00,10.00)
\bezier{40}(19.00,10.00)(20.41,07.18)(21.82,10.00)
\bezier{40}(21.82,10.00)(23.23,12.82)(24.64,10.00)
\bezier{40}(-9.62,30.00)(-8.31,32.82)(-6.82,30.00)
\bezier{40}(-6.82,30.00)(-5.02,27.18)(-4.02,30.00)
\bezier{40}(-3.82,30.00)(-2.71,32.82)(-1.22,30.00)
\bezier{40}(-1.02,30.00)(0.31,27.18)(1.58,30.00)
\bezier{40}(1.58,30.00)(2.99,32.82)(4.7,30.00)

\put( -9.62,30.00){\line(1, -1){04.00}}
\put( -9.62,30.00){\line(1,1){04.00}}
\put( -9.62,30.00){\line(-1,1){04.00}}
\put( -9.62,30.00){\line(-1,-1){04.00}}

\put( -3,26.00){\makebox(0,0)[lc]{${\bf{q}}$}}
\put(15.00,-02.00){\makebox(0,0)[cb]{$d$}}
\end{picture}
\begin{picture}(50.00,45.00)(0.00,-5.00)
%
%
\put(24.60,05.00){\line(0,1){30.00}}
\put(25.40,05.00){\line(0,1){30.00}}
\put(25.00,30.00){\circle*{1.00}}
\put(45.10,10.00){\circle*{1.00}}
\put(25.00,01.00){\makebox(0,0)[cb]{$1s $}}
\put(25.00,39.00){\makebox(0,0)[ct]{$1s$}}
\put(26.00,20.00){\makebox(0,0)[lc]{$$}}
\put(44.60,05.00){\line(0,1){30.00}}
\put(45.40,05.00){\line(0,1){30.00}}
\put(45.00,30.00){\circle*{1.00}}
\put(45.00,01.00){\makebox(0,0)[cb]{$1s$}}
\put(45.00,39.00){\makebox(0,0)[ct]{$e^-$}}
%
%

\bezier{40}(45.64,10.00)(47.05,12.82)(48.46,10.00)
\bezier{40}(48.46,10.00)(49.87,07.18)(51.28,10.00)
\bezier{40}(51.28,10.00)(52.69,12.82)(54.00,10.00)
\bezier{40}(54.00,10.00)(55.41,07.18)(56.82,10.00)
\bezier{40}(56.82,10.00)(58.23,12.82)(59.64,10.00)

\bezier{40}(25.00,30.00)(26.41,32.82)(27.82,30.00)
\bezier{40}(27.82,30.00)(29.23,27.18)(30.64,30.00)
\bezier{40}(30.64,30.00)(32.05,32.82)(33.46,30.00)
\bezier{40}(33.46,30.00)(34.87,27.18)(36.28,30.00)
\bezier{40}(36.28,30.00)(37.69,32.82)(39.00,30.00)
\bezier{40}(39.00,30.00)(40.41,27.18)(41.82,30.00)
\bezier{40}(41.82,30.00)(43.23,32.82)(44.64,30.00)
\put(59.64,10.00){\line(-1, -1){04.00}}
\put(59.64,10.00){\line(-1,1){04.00}}
\put(59.64,10.00){\line(1, 1){04.00}}
\put(59.64,10.00){\line(1,-1){04.00}}
\put( 50.00,15.00){\makebox(0,0)[lc]{${\bf{q}}$}}
\put(35.00,-02.00){\makebox(0,0)[cb]{$e$}}
\end{picture}
\end{center}

\begin{figure}[h]
\begin{minipage}{40pc}
\caption{ Electron loss from the ground state of a helium-like HCI by a single interaction with an external field. The Feynman graphs correspond to the zeroth (\textit{a}) and first (\textit{b}-\textit{e}) orders of the perturbation theory with respect to the electron-electron interaction. The vector ${\bf{q}}$ denotes the momentum transferred by the external field to the HCI. }
\label{fig2}
\end{minipage}\hspace{2pc}%
\end{figure}

\vspace{10.0mm}

\begin{figure}[h]
\begin{minipage}{40pc}
\includegraphics[width=40pc]{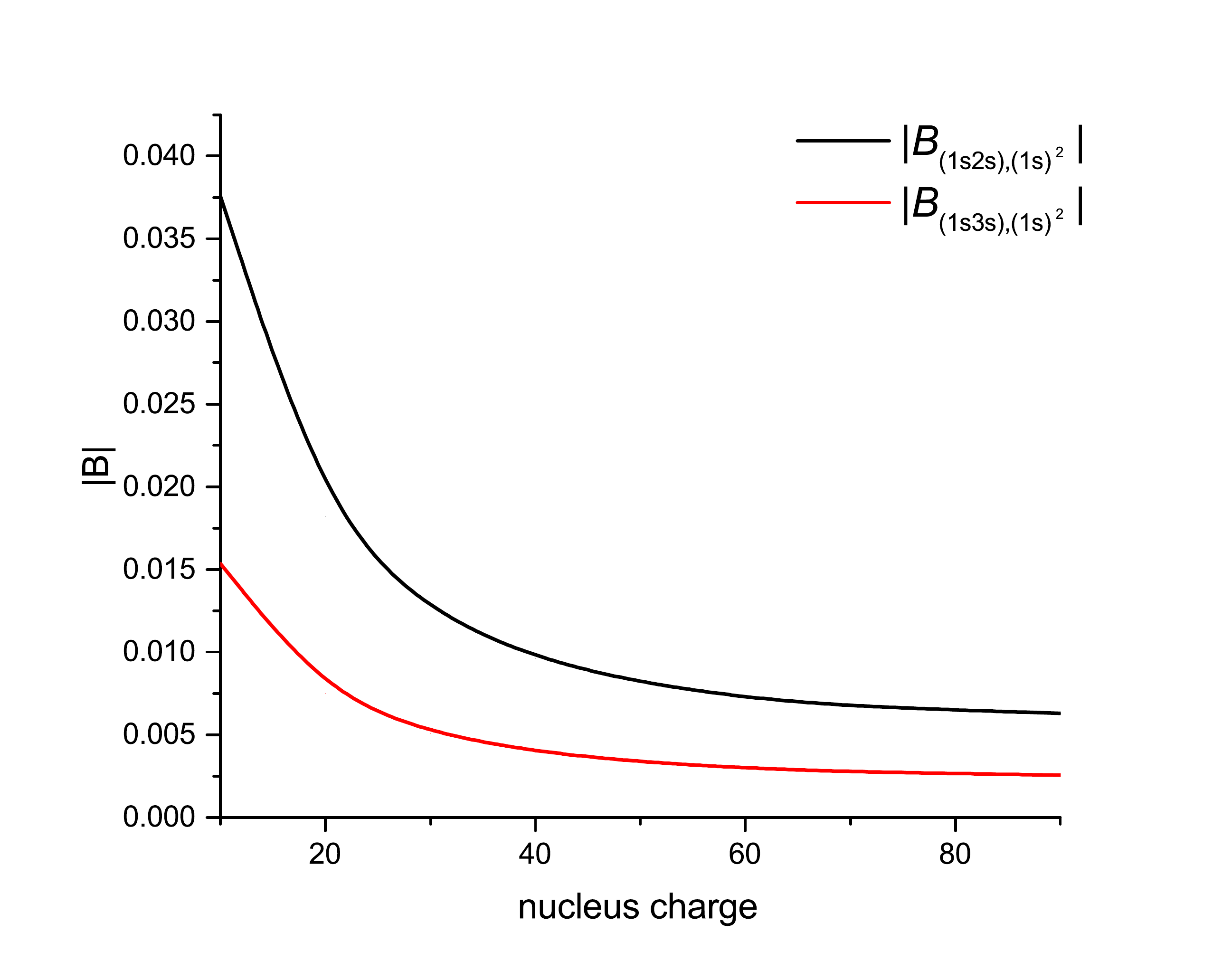}
\caption{ The absolute values of the mixing coefficients for the two-electron configuration $1s^2$ (see \Eq{mix_coef}) given as a function of the nucleus charge of the helium-like ion. }
\label{fig3}
\end{minipage}\hspace{2pc}%
\end{figure}

\begin{figure}[h]
\begin{minipage}{40pc}
\includegraphics[width=40pc]{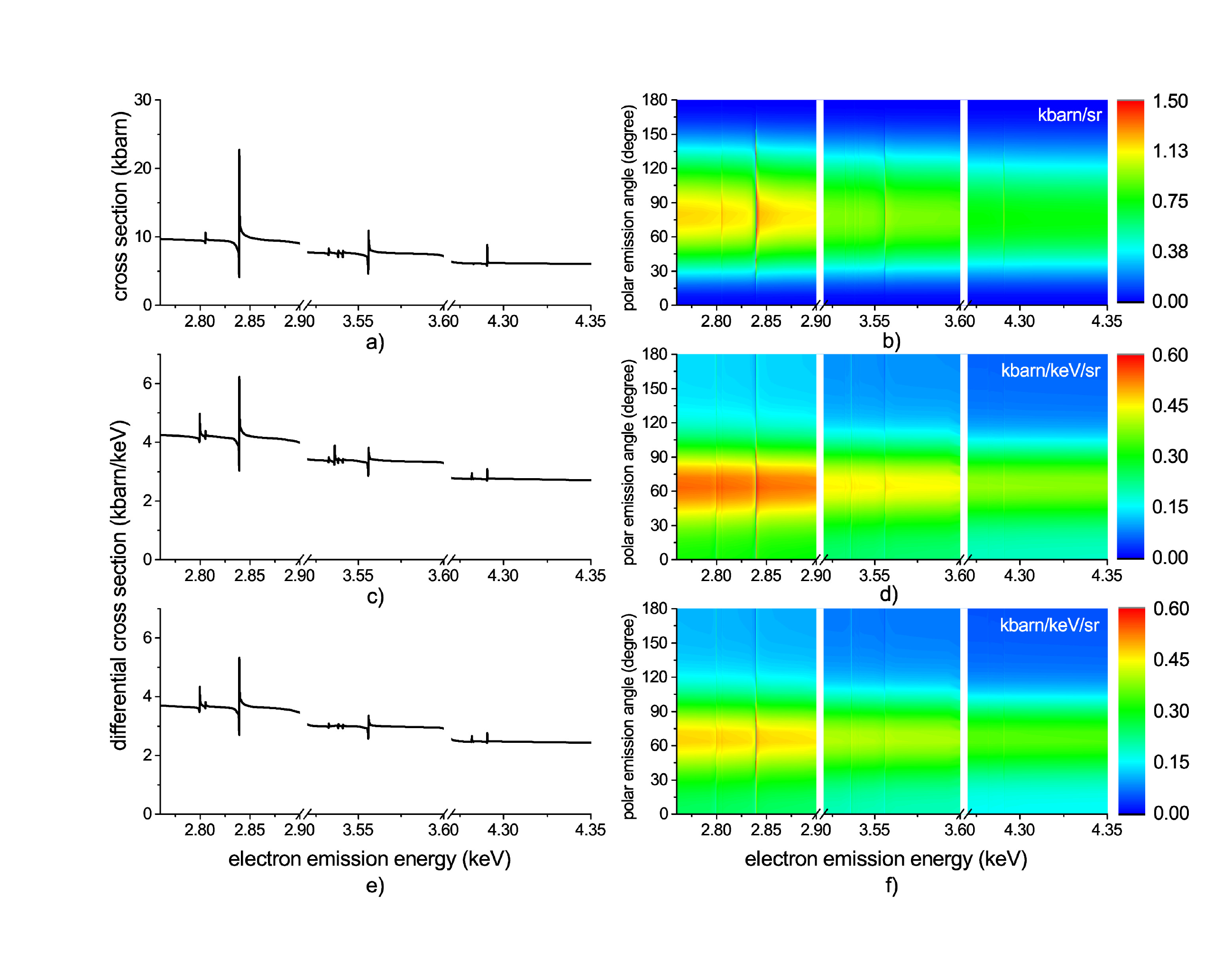}
\caption{ Cross sections for single electron loss from Ca$^{18+}$ $(1s^2)$ by photo absorption (a, b), by the impact of $93$ MeV/u bare nitrogen nucleus (c, d) and neutral nitrogen atom (e, f). The left and right columns present the energy and energy-angular distributions, respectively, of the emitted electrons. The cross sections are given in the rest frame of the HCI. }
\label{fig4}
\end{minipage}\hspace{2pc}%
\end{figure}

\begin{figure}[h]
\begin{minipage}{40pc}
\includegraphics[width=40pc]{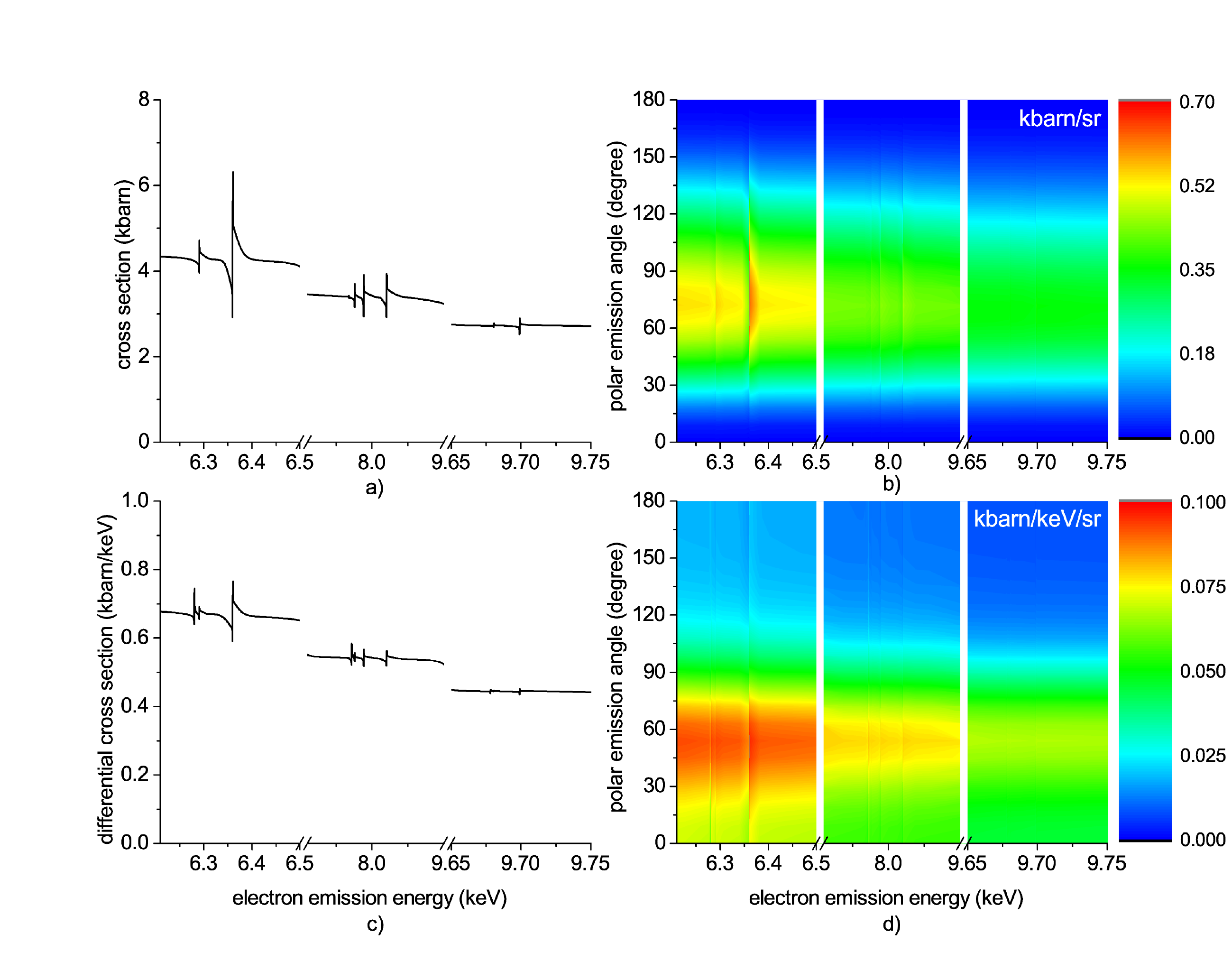}
\caption{ Same as in the corresponding panels ( a - d ) of Fig \ref{fig4}, but for Zn$^{28+}$ $(1s^2)$.}
\label{fig5}
\end{minipage}\hspace{2pc}%
\end{figure}

\begin{figure}[h]
\begin{minipage}{40pc}
\includegraphics[width=40pc]{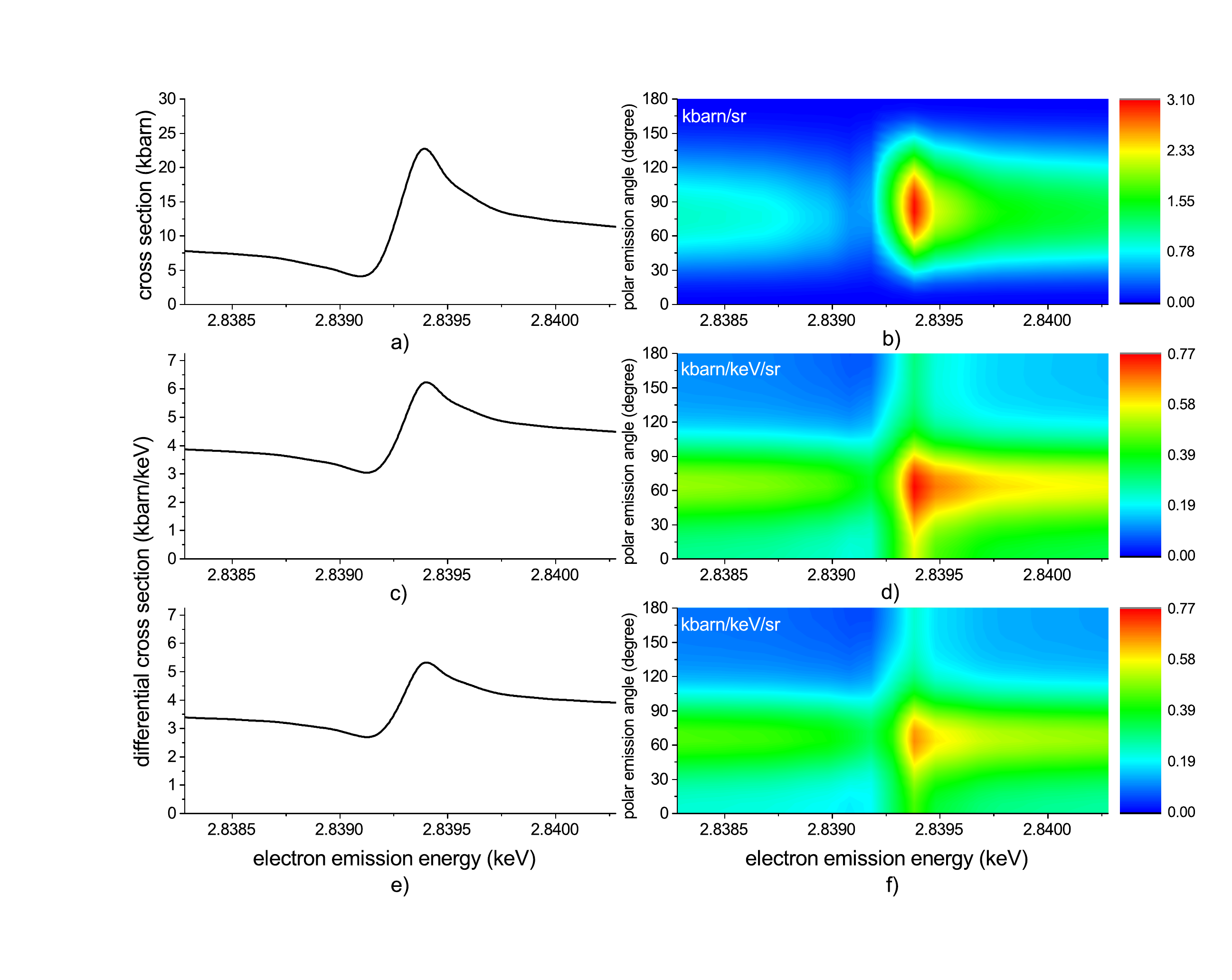}
\caption{ Same as in Fig. \ref{fig4}, but only for the energy range corresponding to the participation of the $(2s2p_{3/2})_{J=1}$ autoionizing state. }
\label{fig6}
\end{minipage}\hspace{2pc}%
\end{figure}

\begin{figure}[h]
\begin{minipage}{40pc}
\includegraphics[width=40pc]{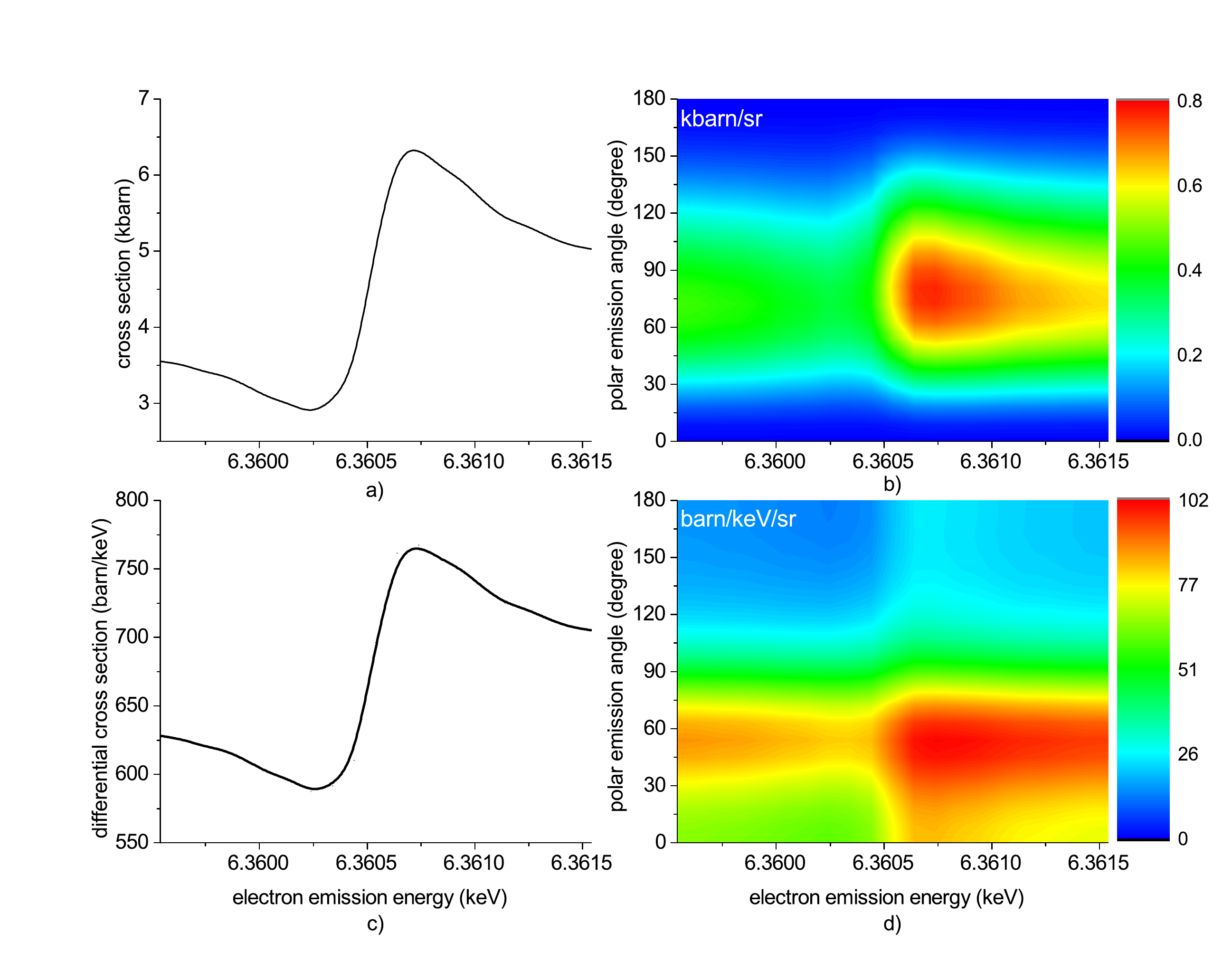}
\caption{ Same as in Fig. \ref{fig6}, but for electron loss from Zn$^{28+}$ $(1s^2)$. }
\label{fig7}
\end{minipage}\hspace{2pc}%
\end{figure}

\begin{figure}[h]
\begin{minipage}{40pc}
\includegraphics[width=40pc]{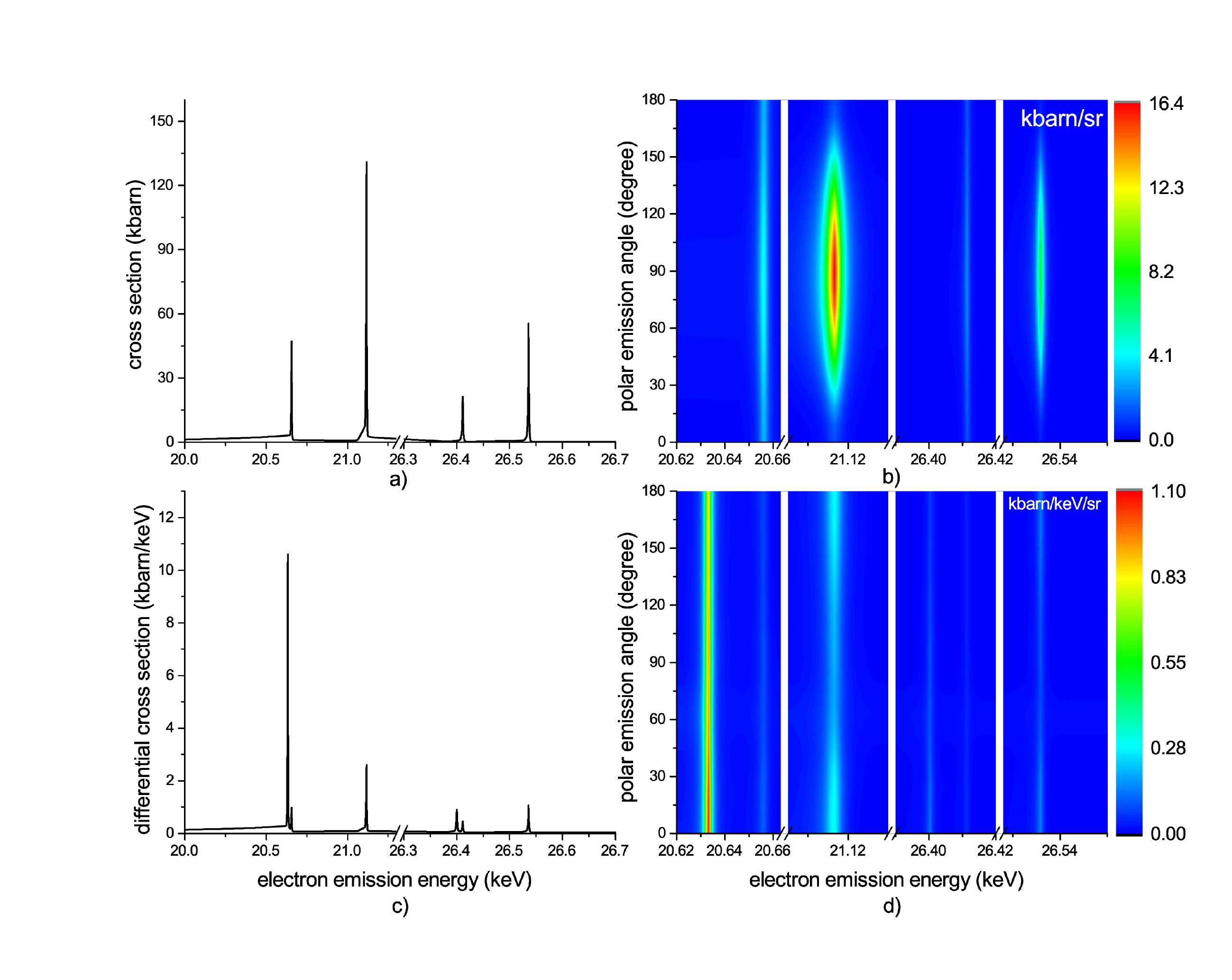}
\caption{ Cross sections for single electron loss from
Xe$^{52+}$ $(1s2s)_{J=0}$ by photo absorption (a, b)
and by the impact of $93$ MeV/u nitrogen (c, d).
The left and right columns present the energy and energy-angular distributions, respectively, of the emitted electrons.
The cross sections are given in the rest frame of the HCI.}
\label{fig8}
\end{minipage}\hspace{2pc}%
\end{figure}

\begin{figure}[h]
\begin{minipage}{40pc}
\includegraphics[width=40pc]{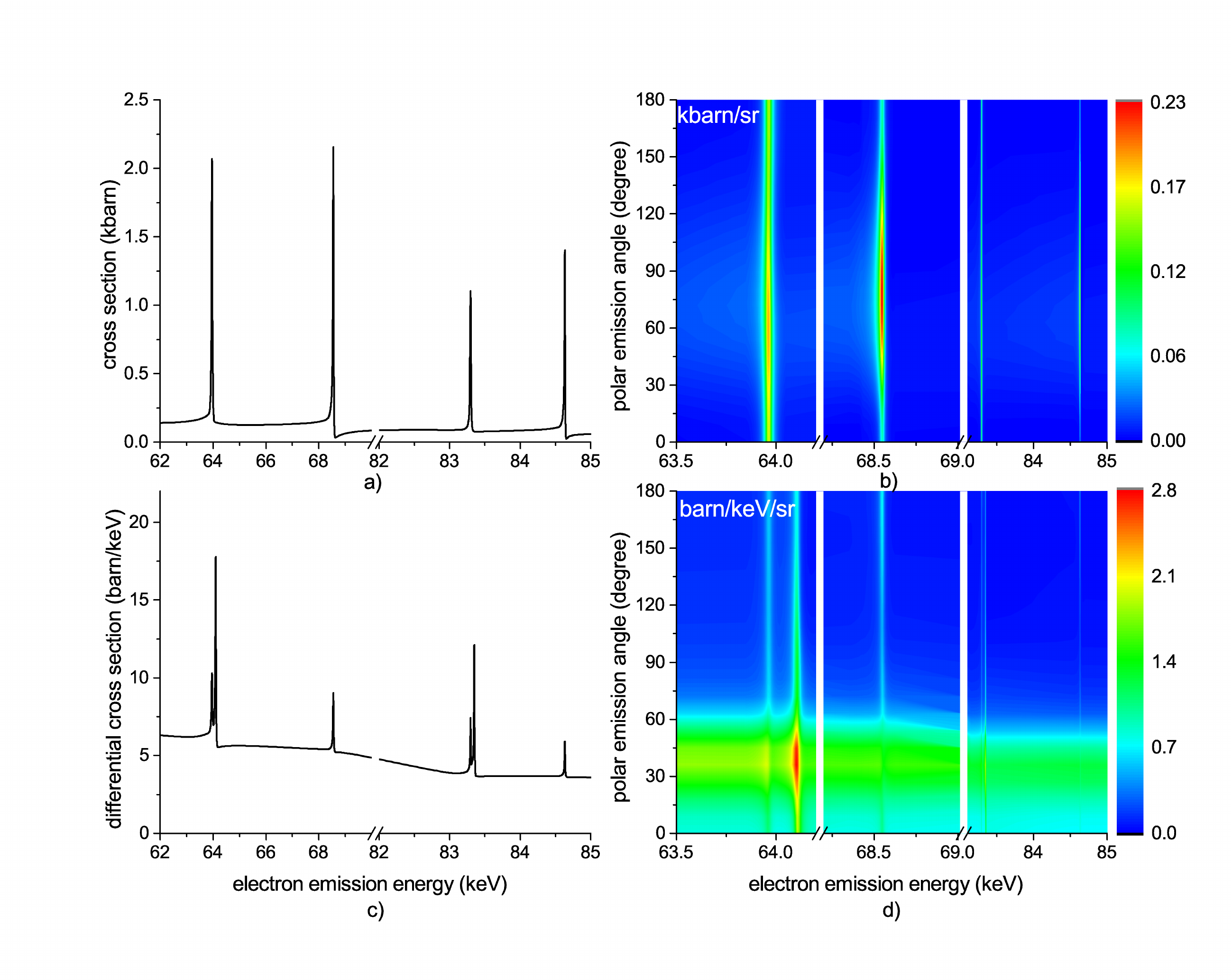}
\caption{Same as in Fig \ref{fig8},
but for U$^{90+}$ $(1s2s)_{J=0}$.}
\label{fig9}
\end{minipage}\hspace{2pc}%
\end{figure}

\begin{figure}[h]
\begin{minipage}{40pc}
\includegraphics[width=40pc]{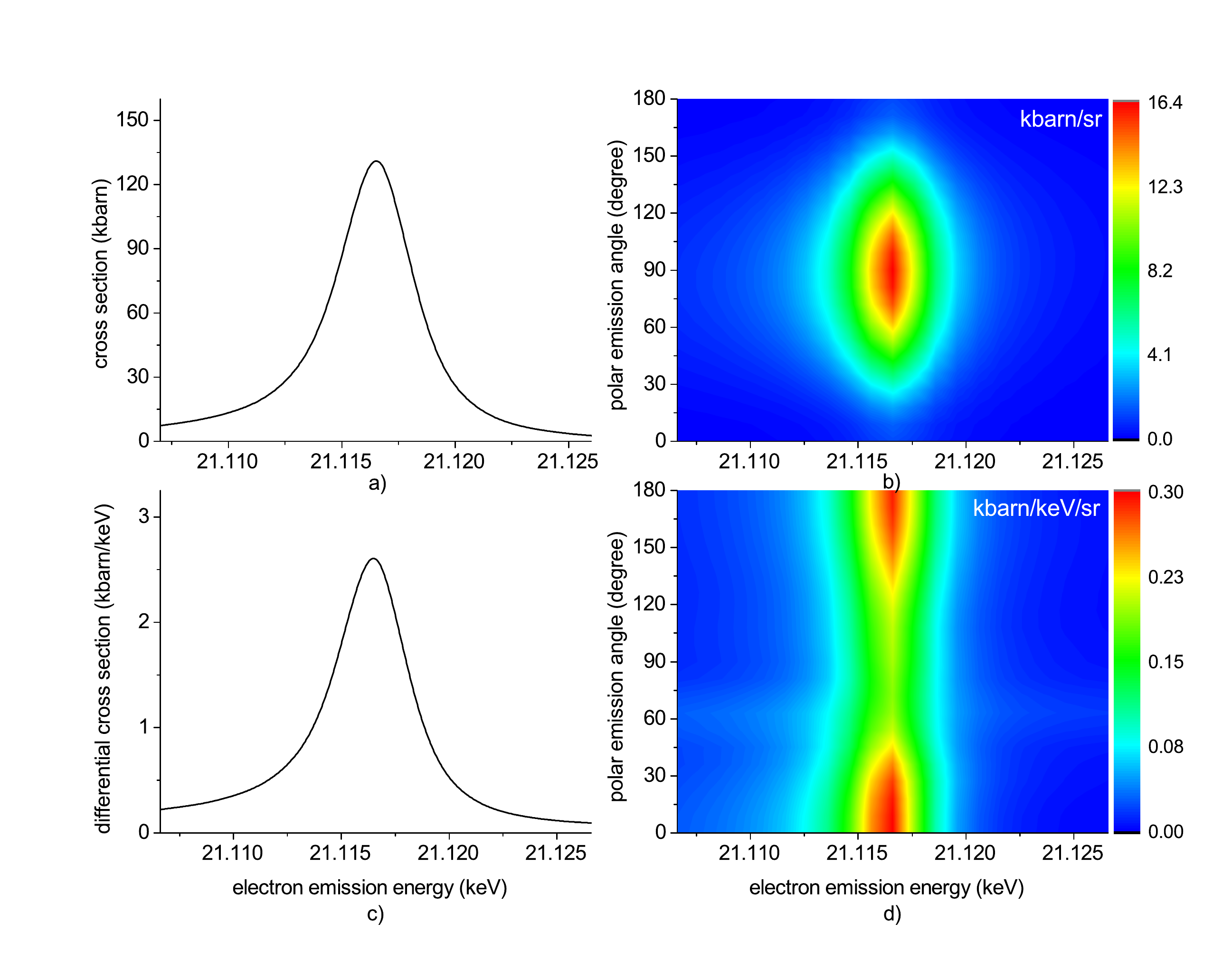}
\caption{ Same as in Fig. \ref{fig8}, but only for
the energy range corresponding to the participation
of the $(2s2p_{3/2})_{J=1}$ autoionizing state. }
\label{fig10}
\end{minipage}\hspace{2pc}%
\end{figure}

\begin{figure}[h]
\begin{minipage}{40pc}
\includegraphics[width=40pc]{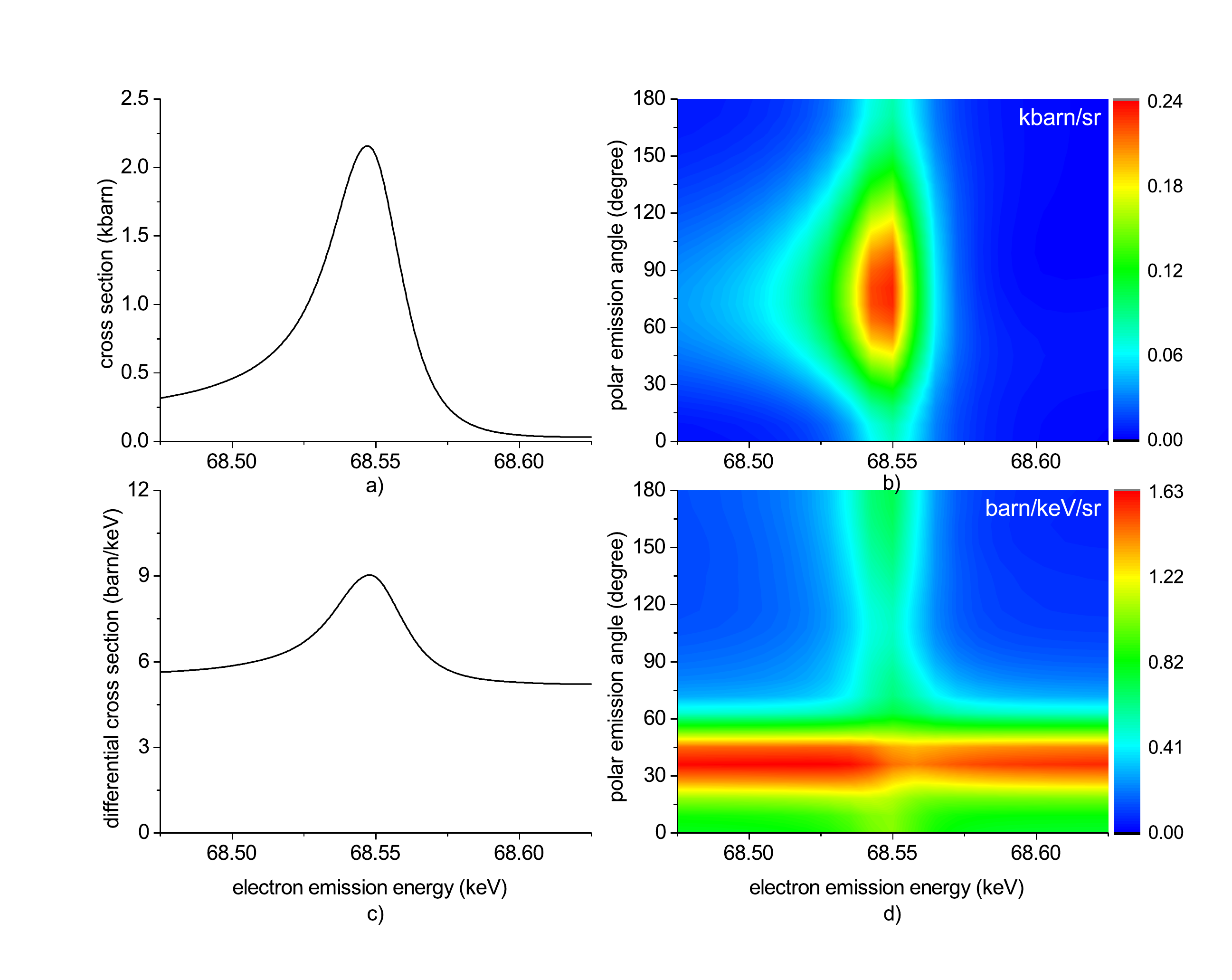}
\caption{ Same as in Fig \ref{fig10}, but for U$^{90+}$ $(1s2s)_{J=0}$. }
\label{fig11}
\end{minipage}\hspace{2pc}%
\end{figure}

\begin{figure}[h]
\begin{minipage}{40pc}
\includegraphics[width=40pc]{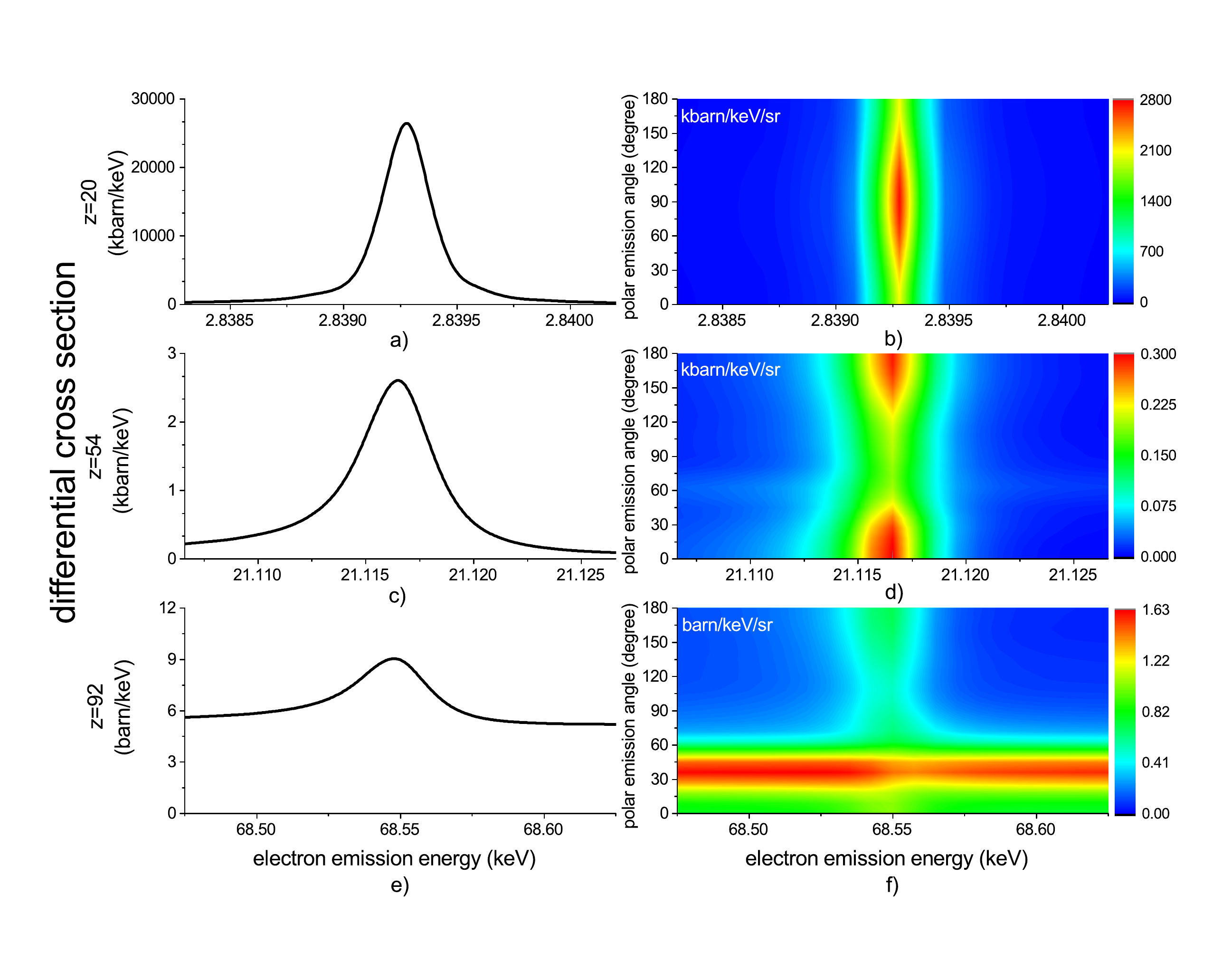}
\caption{ Electron loss from Ca$^{18+}$, Xe$^{52+}$ and U$^{90+}$ ions initially in the $(1s \, 2s)_{J=0}$ state colliding with nitrogen
at $93$ MeV/u impact energy. The spectra are given in the rest frame of the HCIs. The emission energy range corresponds to the participation of
the $(2s \, 2p_{3/2})_{J=1}$ autoionizing state. }
\label{fig12}
\end{minipage}\hspace{2pc}%
\end{figure}

\begin{figure}[h]
\begin{minipage}{40pc}
\includegraphics[width=40pc]{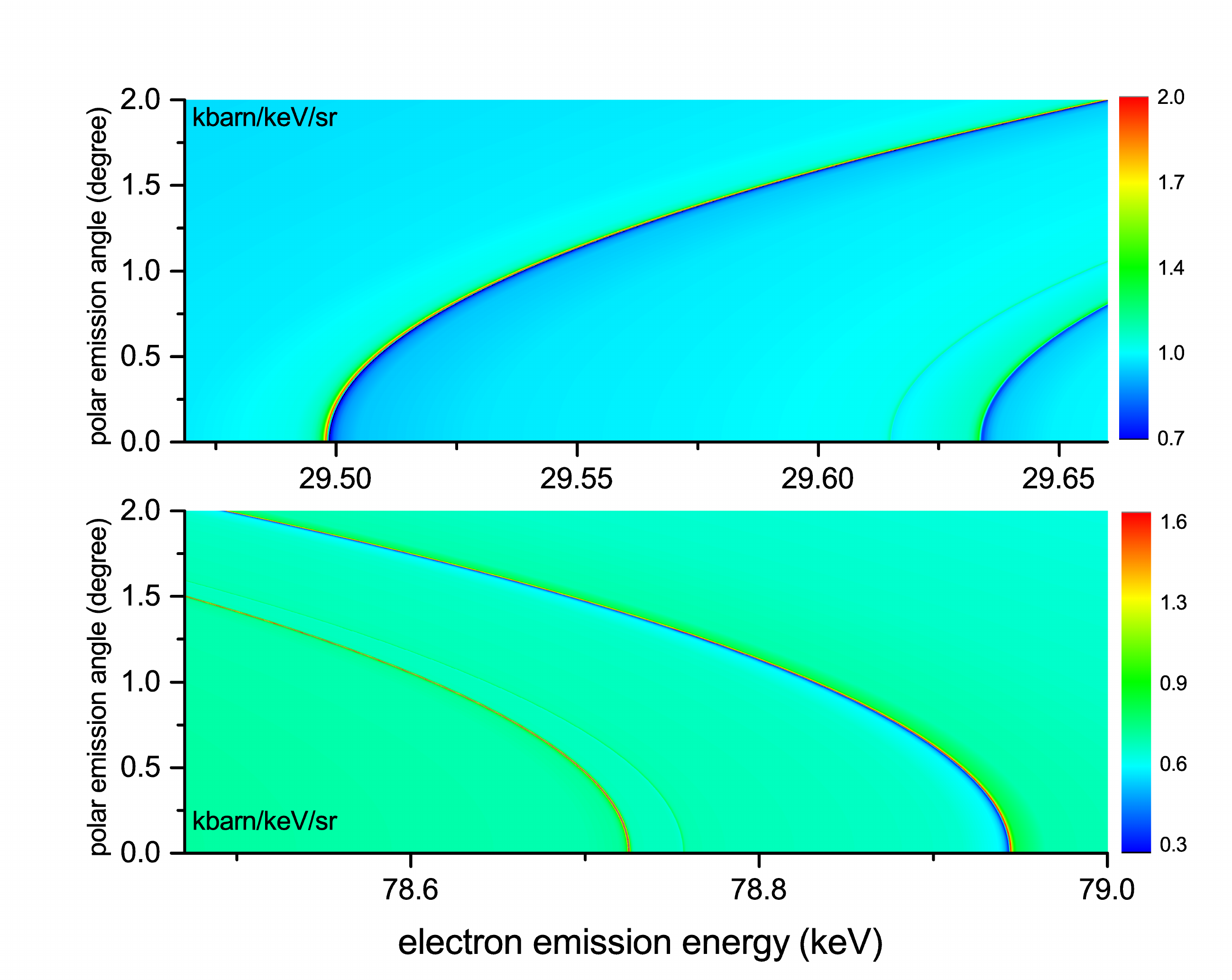}
\caption{ Electron loss from $93$ MeV/u Ca$^{18+}(1s^2)$ colliding with nitrogen. The emission energy range corresponds
to participation of the $(2s^2)$, $(2s \, 2p_{1/2})_{J=1}$ and
$(2s \, 2p_{3/2})_{J=1}$ autoionizing states whose signatures appear
as curved lines (lines corresponding to the above states cross
the zero angle approximately at $29.63$ ($78.72$) keV,
$29.61$ ($78.76$) keV and $29.5$ ($78.94$) keV, respectively). }
\label{fig13}
\end{minipage}\hspace{2pc}%
\end{figure}

\begin{figure}[h]
\begin{minipage}{40pc}
\includegraphics[width=40pc]{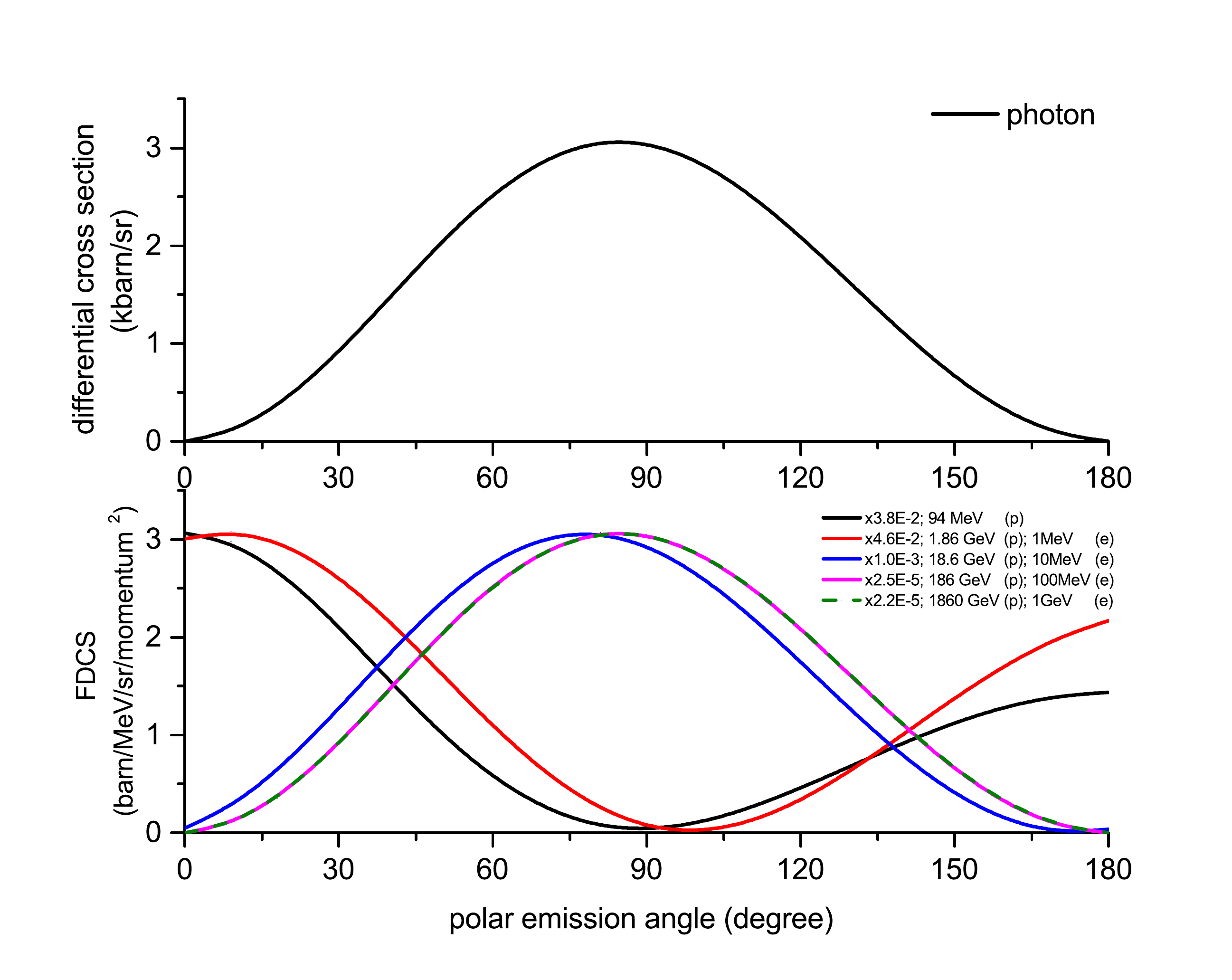}
\caption{ Comparison of electron loss from
Ca$^{18+}(1s^2)$ via photo absorption and
by the impact of a charged particle (proton, electron).
The upper panel shows the angular distribution of
the emitted electrons via absorption of a real photon with an
energy $\hbar \omega \approx 8.311 $ keV
incident along the $z$-axis ($\vartheta = 0^\circ$).
The lower panel displays the fully differential cross section for electron loss 
by absorption of a virtual photon with the same energy $\hbar \omega$ and  momentum $ {\bf q} = ( { \bf q }_{\perp}; q_{min} = \hbar \omega/v) =
( q_{\perp}; 0 ; q_{min} = \hbar \omega/v) $,
where the magnitude of $ q_{\perp}$ is fixed at $ 0.02 \, q_{min} $,
which represents the field of a charged particle incident along
the $z$-axis. In both panels the emitted electron moves in the $(x,z)$-plane. In the lower panel the energy of the incident particle is
given where (p) and (e) denote an equivelocity proton and electron, respectively.
The emission energy corresponds to the resonance with
the $(2s2p_{3/2})_{J=1}$-autoionizing state. In order to be on the same scale, the results for electron loss by charged particles
have been multiplied by corresponding numbers shown
in the lower panel. }
\label{fig14}
\end{minipage}\hspace{2pc}%
\end{figure}

\begin{figure}[h]
\begin{minipage}{40pc}
\includegraphics[width=40pc]{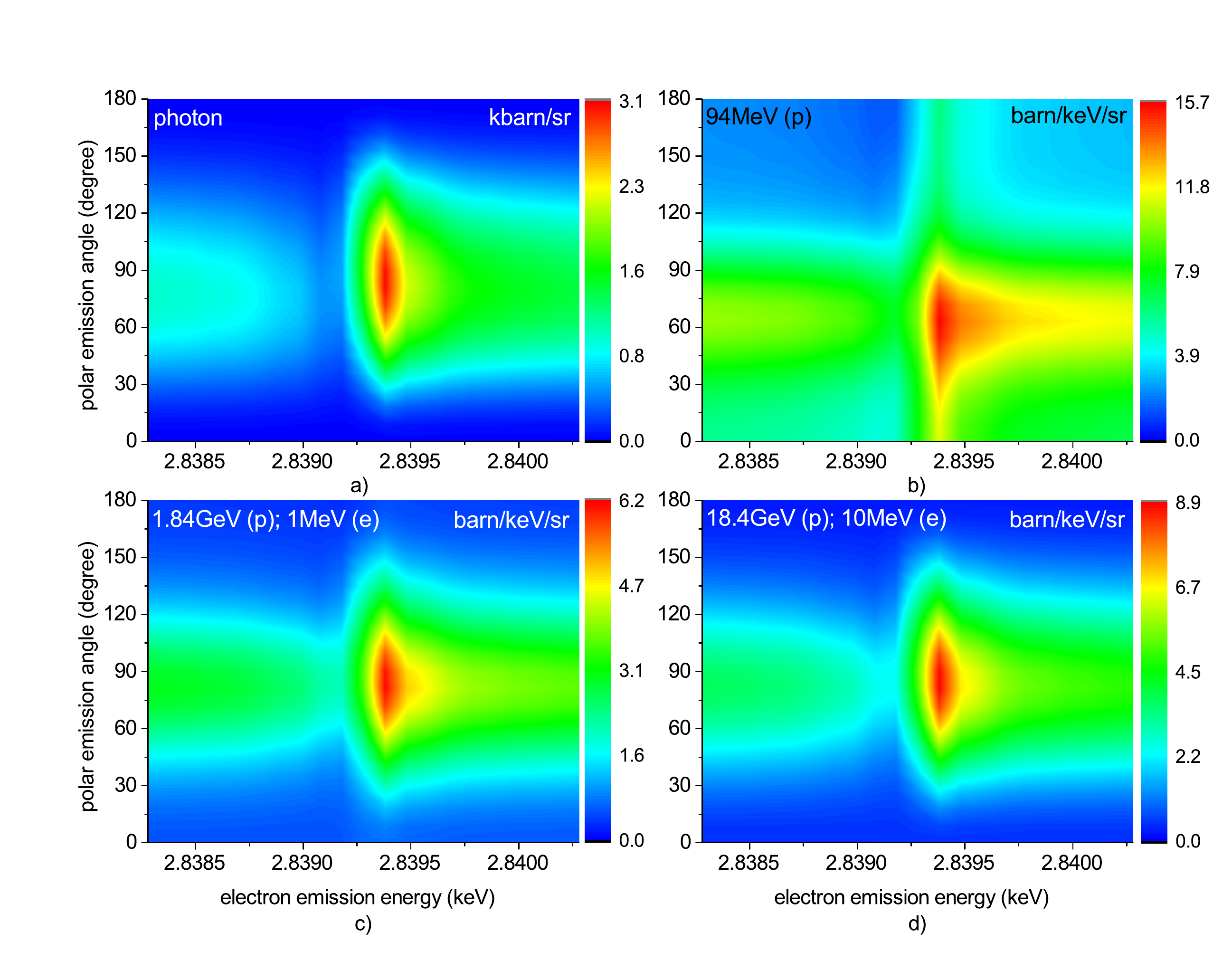}
\caption{ Comparison of electron loss from
Ca$^{18+}(1s^2)$ via photo absorption (a) and
by the impact of a charged particle (proton: b) - d), electron: c) - (d)). Both the incident photon and the charged particle move along the $z$-axis ($\vartheta=0^\circ$).
The emission energy range corresponds to the participation of
the $(2s2p_{3/2})_{J=1}$ autoionizing state. }
\label{fig15}
\end{minipage}\hspace{2pc}%
\end{figure}

\end{document}